\documentclass[11pt,a4paper]{article}
\pdfoutput=1

\usepackage{jheppub}
\usepackage{latexsym}
\usepackage{revsymb}
\usepackage{multirow}
\usepackage{color}
\usepackage[usenames,dvipsnames,svgnames,table]{xcolor}

\usepackage{graphicx}
\usepackage{epsfig}  
\usepackage{epsf}    
\usepackage{dcolumn}
\usepackage{bm}
\usepackage{dcolumn}
\usepackage{textcomp}
\usepackage{float}
\usepackage{subfig}
\usepackage{lineno}
\usepackage{hypcap}
\usepackage[]{hyperref}

\hypersetup{
  bookmarks=true,         
  unicode=false,          
  pdftoolbar=true,        
  pdfmenubar=true,        
  pdffitwindow=true,     
  pdfstartview={FitH},    
  pdfsubject={Neutrino Oscillations Phenomenology},   
  pdfnewwindow=true,      
  pdfcreator={RevTeX},
  colorlinks=true,       
  linkcolor=red,          
  citecolor=blue,        
  filecolor=black,      
  urlcolor=blue,           
}

\newcommand{\beq}{\begin{equation}}
\newcommand{\eeq}{\end{equation}}
\newcommand{\beqa}{\begin{eqnarray}}
\newcommand{\eeqa}{\end{eqnarray}}

\newcommand{\ty}{{\theta_{13}}}
\newcommand{\tz}{{\theta_{23}}}

\begin{document}
\title{Probing CP violation with T2K, NO$\nu$A and DUNE in the presence of non-unitarity}
\author[a]{Debajyoti Dutta,}
\author[b]{Pomita Ghoshal}

\affiliation[a]{Harish-Chandra Research Institute, Chhatnag Road, Jhunsi,
Allahabad 211019, India}
\affiliation[b]{Department of Physics, LNM Institute of Information Technology (LNMIIT),\\
Rupa-ki-Nangal, post-Sumel, via-Jamdoli, Jaipur-302 031, Rajasthan, India}

\emailAdd{debajyotidutta@hri.res.in}
\emailAdd{pomita.ghoshal@gmail.com}

\begin{abstract}
{
The presence of non-unitary neutrino mixing can affect the measurement of the three-neutrino leptonic Dirac CP phase
and hamper efforts to probe CP violation due to degeneracies of the extra non-unitary CP phase with the standard CP phase.
We study the effect of including non-unitarity on probing CP violation with the long-baseline
experiments NO$\nu$A, T2K and DUNE. We analyze the effect of non-unitary mixing 
at the level of oscillation probabilities associated with the relevant baselines, and present the
CP violation sensitivity for the individual experiments and their combination. Our results show that
there is an improvement in the CP violation sensitivity of the combination compared to the individual
experiments.   
}
\end{abstract}

\arxivnumber{1607.02500}

\maketitle

\section{Introduction}
\label{Introduction}
The standard three-neutrino oscillation framework consists of three mixing angles $\theta_{12}$, $\theta_{13}$ and $\theta_{23}$, two mass-squared differences $\Delta m_{31}^2$ and $\Delta m_{21}^2$, and the leptonic Dirac CP phase $\delta_{CP}$. Currently, through a combination of accelerator, reactor, solar and atmospheric neutrino experiments, the standard neutrino mixing parameters have been determined to varying degrees of precision. The leptonic CP phase $\delta_{CP}$ still remains one of the least known parameters in this scheme. The possibility of CP violation has not yet been determined with any significant degree of precision. A value of $\delta_{CP}$ different from 0 or 180$^o$ would indicate CP violation in the lepton sector. In the quark sector, CP violation has been observed and can be explained by the complex phase in the CKM mixing matrix due to complex Yukawa couplings or complex Higgs field vacuum expectation values \cite{1, 2}. Similar origins for the leptonic CP phase have been proposed, but are yet to be experimentally verified. 

The possibility of leptonic CP violation is an important issue, since it can provide an explanation for the observed matter-antimatter asymmetry in the universe through leptogenesis \cite{3, 4}. Also, a knowledge of the CP phase would complete the picture of the PMNS \cite{5, 6, 7} mixing matrix. There have been several attempts to study CP sensitivity using current and proposed long-baseline experiments - with conventional superbeams \cite{8, 9, 10, 11, 12}, long-baseline experiments like NO$\nu$A \cite{13} and T2K \cite{14}, future experiments like DUNE \cite{15, 16}, LBNO \cite{17} and T2HK \cite{18} etc. The effect of a large reactor mixing angle $\theta_{13}$ and the lack of knowledge of the neutrino mass hierarchy has been analyzed in detail. The new experimental set-ups offer great promise in providing enough statistics and mutual synergies in order to resolve existing degeneracies and determine these unknowns in current neutrino physics. 

However, there is an additional complication which may hamper the determination of leptonic CP violation - the possible presence of new physics which can give rise to additional CP phases which mimic the leptonic CP phase and lead to further degeneracies. Recently, this issue has been analyzed in \cite{19} in the context of a light sterile neutrino in the DUNE experiment. Non-unitary neutrino mixing can also give an extra CP violating phase which results in this kind of degeneracy. Focusing on non-standard interactions and a four-neutrino scenario, CP-invariant new physics at DUNE is studied in \cite{20}.

Non-unitarity (NU) in the neutrino mixing matrix is one of the possible departures from the standard three-neutrino mixing framework, and can occur due to the induction of neutrino mass through the type-I seesaw mechanism. If the messenger fermions involved are within the reach of the Large Hadron Collider, a rectangular leptonic mixing matrix would be obtained, giving an effectively non-unitary neutrino mixing matrix \cite{ 21, 22, 23, 24, 25}. Such a framework has a new non-unitary phase which is degenerate with the standard CP phase and leads to a change in the sensitivity to CP violation. In \cite{26}, this degeneracy is discussed at the level of oscillation probabilities, and in \cite{27} a solution in offered in terms of the upgrade of T2HK to TNT2K. The effect of NU is also studied at the probability level in vacuum for the T2K, NOVA and DUNE experiments in \cite{28}. IN \cite{29}, unitarity of the PMNS matrix is tested using direct and indirect method.

Measurement of CP violation is one of the most important question in neutrino sector and these long baseline experiments are able to probe it. CP violation sensitivity measurement at DUNE in different contest can be found in \cite{30, 31, 32, 33}. Similar studies for T2K and NO$\nu$A can be found in \cite{34, 35, 36, 37}. Capability to measure CP violation in these experiments with non standard interactions can be found in \cite{38, 39, 40, 41}.
 
 In the present work, we focus on the effect of non-unitarity on probing CP violation with the long-baseline
experiments NO$\nu$A, T2K and DUNE. We analyze the effect of non-unitary mixing on the oscillation probabilities associated with the given experiment baselines, and describe the CP violation sensitivity for the individual experiments and their combination. 
We show that there is 
an improvement in the CP violation sensitivity of the combination.    
The analysis is performed with realistic simulations of all three experiments using the standard simulator GLoBES \cite{42, 43},
which includes matter effects in the oscillation probability as well as relevant systematics for each experiment. We have incorporated MonteCUBES's \cite{44} Non Unitarity Engine (NUE) with GLoBES while performing this analysis.

The paper is organized as follows: in Section II we discuss the oscillation probability $P (\nu_{\mu} \rightarrow \nu_{e})$ relevant to the experiments considered in the presence of non-unitary mixing, explaining the parameterization adopted. Section III describes the technical details of the experiments NO$\nu$A, T2K and DUNE, and outlines the simulation procedure followed by us to compute the CP violation sensitivity for the experiments and their combination. In Section IV, we present figures of the probability $P (\nu_{\mu} \rightarrow \nu_{e})$ as a function of neutrino energy for the relevant baselines, which show how the probability gets undetermined for specific values of the CP phase $\delta_{CP}$ when non-unitarity is included. The CP asymmetry is also depicted for different cases. Section V gives the results for the event numbers and the CP violation sensitivity of the different experiments. We conclude with a discussion of the results in Section VI. 
 
\section{Neutrino Oscillation Probability with a Neutral Heavy Lepton (NHL)}

The most general structure of the parametrization adopted in this work starts with \cite{45} and the  symmetrical parametrization technique can be found in \cite{46}. In the presence of a Neutral Heavy Lepton, the $3\times 3$ neutino mixing matrix is no longer unitary and gets modified as  \begin{equation}N = N^{NP}U, \end{equation} where U is the $3\times 3$ PMNS matrix.  $N^{NP}$ is the left triangular matrix and can be written as \cite{24}
 
 \begin{equation}
N^{NP} = 
\begin{pmatrix}
\alpha_{11} & 0 & 0 \\
\alpha_{21} & \alpha_{22} & 0 \\
\alpha_{31} & \alpha_{32} & \alpha_{33} \\
\end{pmatrix}
\end{equation}

Due to this structure of the pre factor matrix, there remain only four extra parameters which affect the neutrino oscillations - the real parameters $\alpha_{11}$ and $\alpha_{22}$, one complex parameter $|\alpha_{21}|$ and the phase associated with $|\alpha_{21}|$. In the presence of the non unitary mixing matrix, the electron neutrino appearance probability changes in vacuum, as explained in \cite{24, 27}. The expression for $P_{\mu e}$ in the presence of NU can be written as
\begin{equation}
P_{\mu e} = (\alpha_{11}\alpha_{22})^2 P^{3\times 3}_{\mu e}+\alpha_{11}^2\alpha_{22}|\alpha_{21}|P^{I}_{\mu e}+\alpha_{11}^2|\alpha_{21}|^2.
\end{equation} 
 Here, $P^{3\times 3}_{\mu e}$ is the standard three flavor neutrino oscillation probability and $P^{I}_{\mu e}$ is the oscillation probability containing the extra phase that appears due to the non unitary nature of the mixing matrix. $P^{3\times 3}_{\mu e}$ in the above expression can be written as :
 \begin{equation}
 \begin{split}
 P^{3\times 3}_{\mu e} &= 4[\cos^2\theta_{12}\:\cos^2\theta_{23}\:\sin^2\theta_{12}\:\sin^2(\frac{\bigtriangleup m^2_{21}L}{4E_{\nu}})+\cos^2\theta_{13}\:\sin^2\theta_{13}\:\sin^2\theta_{23}\:\sin^2(\frac{\bigtriangleup m^2_{31}L}{4E_{\nu}})] \\
 & +\sin(2\theta_{12})\:\sin\theta_{13}\:\sin(2\theta_{23})\:\sin(\frac{\bigtriangleup m^2_{21}L}{2E_{\nu}})\:\sin(\frac{\bigtriangleup m^2_{31}L}{4E_{\nu}})\:\cos(\frac{\bigtriangleup m^2_{31}L}{4E_{\nu}}-I_{123}).
 \end{split}
  \end{equation}
 
 And
  \begin{equation}
 \begin{split}
 P^{I}_{\mu e} & = -2[\sin(2\theta_{13})\:\sin\theta_{23}\:\sin(\frac{\bigtriangleup m^2_{31}L}{4E_{\nu}})\:\sin(\frac{\bigtriangleup m^2_{31}L}{4E_{\nu}}+\phi_{NP}-I_{123})] \\
 &-\cos\theta_{13}\:\cos\theta_{23}\:\sin(2\theta_{12})\:\sin(\frac{\bigtriangleup m^2_{21}L}{2E_{\nu}})\sin({\phi_{NP}}),
  \end{split}
  \end{equation}
   where
 $I_{123} = -\delta_{CP} = \phi_{12}-\phi_{13}+\phi_{23}$ and $\phi_{NP} = \phi_{12}-\rm{Arg}(\alpha_{21})$ and $\alpha_{21} = |\alpha_{21}|\exp(\phi_{NP})$ .
\section{Details of Experiments}
\label{Experiments}
In this section we summarize the technical details of all the three superbeam experiments to be discussed for the sake of completeness. \\

{\bf{T2K}}

The Tokai to Kamioka (T2K) experiment \cite{14} in Japan is an ongoing neutrino oscillation experiment. The main goals are to observe
$\nu_{\mu}\rightarrow\nu_{e}$ oscillations and to measure $\ty$. The $\nu_{\mu}$ beam generated at the J-PARC accelerator facility is directed to Kamioka where a 22.5 kton 
water \v{C}erenkov detector is placed at a $2.5^\circ$ off-axis angle \cite{47} to discriminate between electron and muon neutrino interactions. The baseline is 295 km and beam power is 750 kW. The first oscillation maximum of the $\nu_\mu\rightarrow\nu_e$ appearance probability occurs at 0.6 GeV, and after that the flux falls off quite rapidly.
With a plan to run for a total exposure of $\sim 8\times10^{21}$ 
protons on target (POT), it has already collected 10\% of data while running in $\nu$ mode. It has its near detector sitting at 280 m away from the source to measure the neutrino flux. We have not considered the presence of the ND in this work. We are also assuming that T2K will run for 3 years in $\nu$ and 3 years in $\bar{\nu}$ mode. Details regarding the detector efficiencies and 
background events used in our work  have been taken from \cite{35}.
 We have assumed 2.5\% (5\%) signal and  20\% (5\%) background normalisation errors in $\nu_{\mu}$ ($\nu_{e}$) signal. 
 
{\bf{NO$\nu$A}}

The NuMI{\footnote{Neutrinos at the Main Injector} }
Off-axis $\nu_{e}$ Appearance experiment (NO$\nu$A) \cite{13} 
is another ongoing neutrino experiment 
in the US. The main goal of this superbeam experiment is to determine the 
octant of $\tz$, the neutrino mass hierarchy, $\ty$ and leptonic CP violation using
$\nu_{\mu}\rightarrow\nu_{e}$ oscillations. A Totally Active 
Scintillator Detector (TASD) of mass 14 kton is placed in Ash River, Minnesota. The detector is positioned on the surface at an off-axis angle of 14 mrad ($0.8^\circ$) and the baseline is 810 km.
This $\nu_{\mu}$ beam peaks at 2 GeV.  
The experiment is scheduled to run for 3 years with a beam power of $0.7$ MW.
A proton beam of 120 GeV delivers $6\times 10^{20}$ 
POT per year. A near detector of mass 290 ton located 1 km away from the NuMI target
is also in off-axis position to measure unoscillated neutrinos. Recently, the NO$\nu$A collaboration has reported their first result with less than 10\% of the planned exposure and the result is consistent with maximal $\tz$ \cite{48}. The details of signal and background
events and the detector efficiencies have been taken from
\cite{35}.

{\bf{DUNE}}

 The Deep Underground Neutrino Experiment (DUNE) is scheduled to come online in $\sim2025$. The experimental specifications are very similar to
LBNE \cite{15, 16}. The United States based DUNE experiment is capable of addressing all the three most important questions in the neutrino sector - mass hierarchy, octant of $\tz$ and existence 
of CP violation in the leptonic sector. The DUNE baseline is so optimized that both determining the neutrino mass hierarchy as well as searching for leptonic CP violation can be simultaneously carried out within the same experiment. The ${\nu_{\mu}(\bar{\nu}_{\mu}})$ super-beam originating at Fermilab will be detected by a 35-40 kt Liquid Argon (LAr) far detector installed at a distance of 1300 km in the Homestake mine in South Dakota. 
 A 1.2 MW - 120 GeV proton beam will deliver $10^{21}$ 
protons-on-target (POT) per year. The experiment plans to run for 10 years both in neutrino and anti-neutrino mode, corresponding  to a total
exposure of $35\times10^{22}$ kt-POT-yr. For DUNE, we have not considered the tau events in the background during this work.

All the experimental details,
such as signal and background definitions as well as the detector efficiencies 
for all the three experiments are tabulated in Table 1.

 \begin{table}[!h]
\begin{center}
\begin{tabular}{|c|c|c|c|c|}
\hline 
Experiment & Signal & Signal  & Background   &   Calibration error   \tabularnewline
&   &    norm error & norm error   &   Signal \qquad	Background  \tabularnewline
\hline 
DUNE & $\nu_e$ & 5\%  & 10\% & 5$\%$ \qquad 5\% \tabularnewline
 
 & $\nu_{\mu}$ & 5\%   & 10\%  & 5$\%$ \qquad 5\% \tabularnewline
\hline 
NO$\nu$A & $\nu_e$ & 5\% & 10\% & 0.01\% \qquad 0.01\% \tabularnewline
 
 & $\nu_{\mu}$ & 2.5\%   & 10\%  & 0.01\% \qquad 0.01\%  \tabularnewline
\hline 
T2K & $\nu_e$ & 5\%  & 5\% & 0.01\% \qquad 0.01\% \tabularnewline
 
 & $\nu_{\mu}$ & 2.5\%   & 20\%  & 0.01\% \qquad 0.01\%  \tabularnewline
\hline 
\end{tabular}
\caption{Systematics uncertainties for T2K, NO$\nu$A and DUNE}
\label{Table 2}

\par\end{center}
\end{table}

 \begin{table}[!h]
\begin{center}
\begin{tabular}{|c|c|c|c|c|c|}
\hline 
Experiment& Signal & Signal &  Energy & Runtime (yrs)    &   Detector Mass (Type)   \tabularnewline
&	& Efficiencies & Resolutions & $\nu + \bar{\nu}$   &   \tabularnewline
\hline 

DUNE & $\nu_{e}^{CC}$&80\% & $0.15/\sqrt{E}$  & 5 + 5 & 35 kton (LArTPC) \tabularnewline
 
 & $\nu_{\mu}^{CC}$& 85\%& $0.20/\sqrt{E}$   &  & \tabularnewline
\hline 
NO$\nu$A &$\nu_{e}^{CC}$ & 55\% & $0.085/\sqrt{E}$ & 3 + 3 & 14 kton (TASD) \tabularnewline
 
 & $\nu_{\mu}^{CC}$& 85\%& $0.06/\sqrt{E}$   &   &  \tabularnewline
\hline 
T2K & $\nu_{e}^{CC}$& 50\% & $0.085/\sqrt{E}$  & 3 + 3 & 22.5 kton (WC) \tabularnewline
 
 & $\nu_{\mu}^{CC}$& 90\%  & $0.085/\sqrt{E}$  &  &  \tabularnewline
\hline 
\end{tabular}
\caption{Simulation details like signal efficiencies, energy resolutions, total exposures and detector mass for T2K, NO$\nu$A and DUNE}
\label{Table 2}

\par\end{center}
\end{table}

{\bf{Simulation Parameters}} 

Throughout this analysis, we have fixed the values of the three-flavor neutrino oscillation parameters except $\delta_{CP}$. The values of the solar and reactor mixing angles are fixed at $\theta_{12} = 33.48^0$ and $\theta_{13} = 8.5^0$ \cite{49} respectively. The $3\sigma$ allowed range of $\theta_{23}$ is [38.3, 53.3] with a best fit value of $42.3^0 (49.5^0)$ assuming NH (IH) as the true hierarchy. The octant of $\theta_{23}$ and hence the best fit value is not yet fixed as seen from different global analyses \cite{50, 51}. We assume the maximal value of the atmospheric mixing angle in this study i.e. $\theta_{23} = 45^0$. 
The solar and atmospheric mass square differences are fixed at $\Delta{m}^2_{21} = 7.5 \times 10^{-5}$ e$V^2$ and $\Delta{m}^2_{31} = 2.457 \times 10^{-3}$ e$V^2$ respectively \cite{35}.

The updated bounds on the non-unitarity parameters are $\alpha_{11}^2 \geq 0.989$, $\alpha_{22}^2 \geq 0.999$ and $|\alpha_{21}|^2 \leq 6.6\times 10^{-4}$ at 90\% C.L. \cite{24, 27}. So while performing the $\chi^2$ analysis, we have assumed their central values as the true values. 

\section{Probability and CP Asymmetry}

In this section, we study the effect of non-unitarity on the oscillation probability $\rm P(\nu_{\mu} (\bar{\nu_{\mu}})\rightarrow \nu_e (\bar{\nu_e}))$ and the CP asymmetry. 

\begin{figure}[t]
\centering
\includegraphics[width=0.49\textwidth]{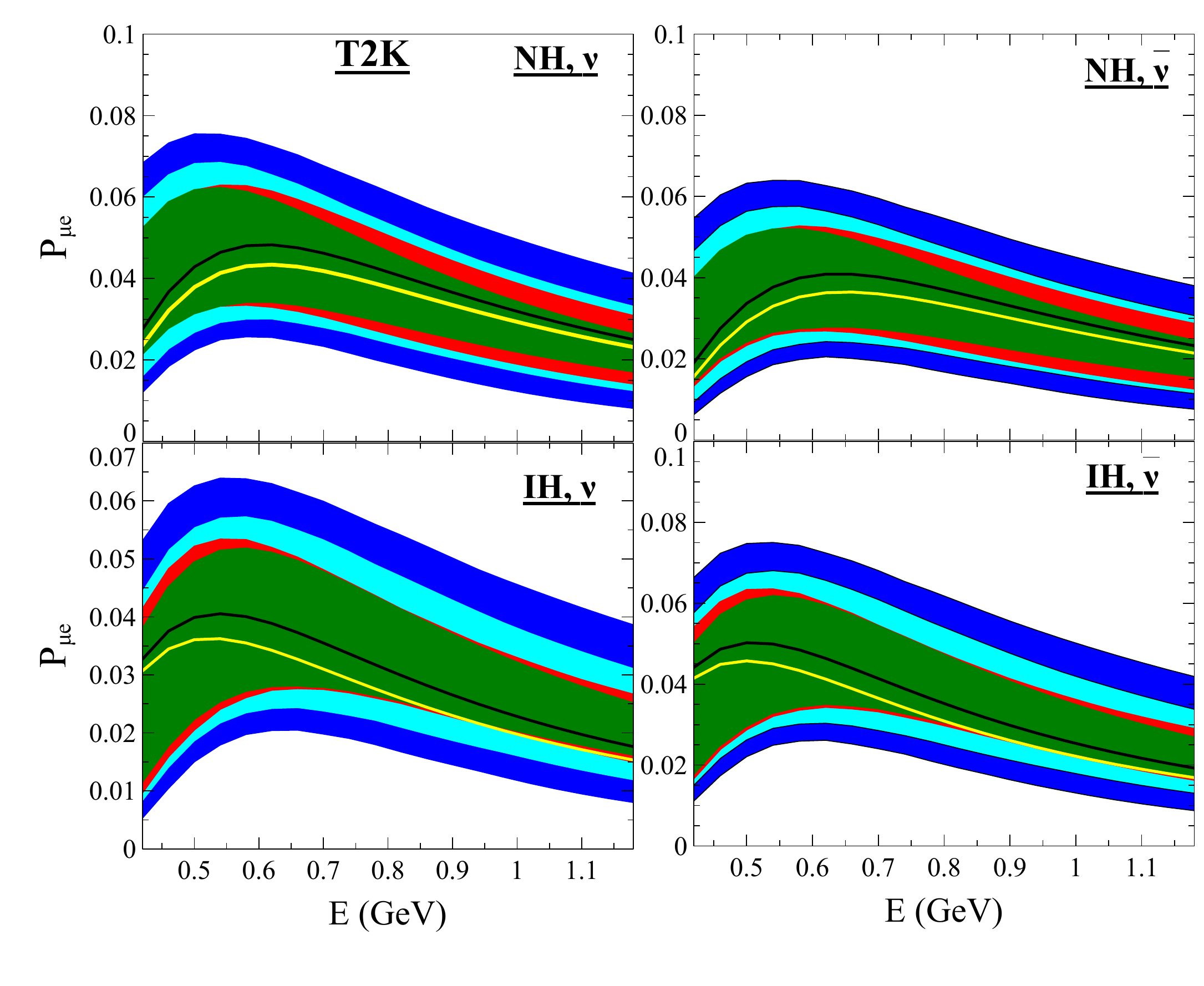}
\includegraphics[width=0.49\textwidth]{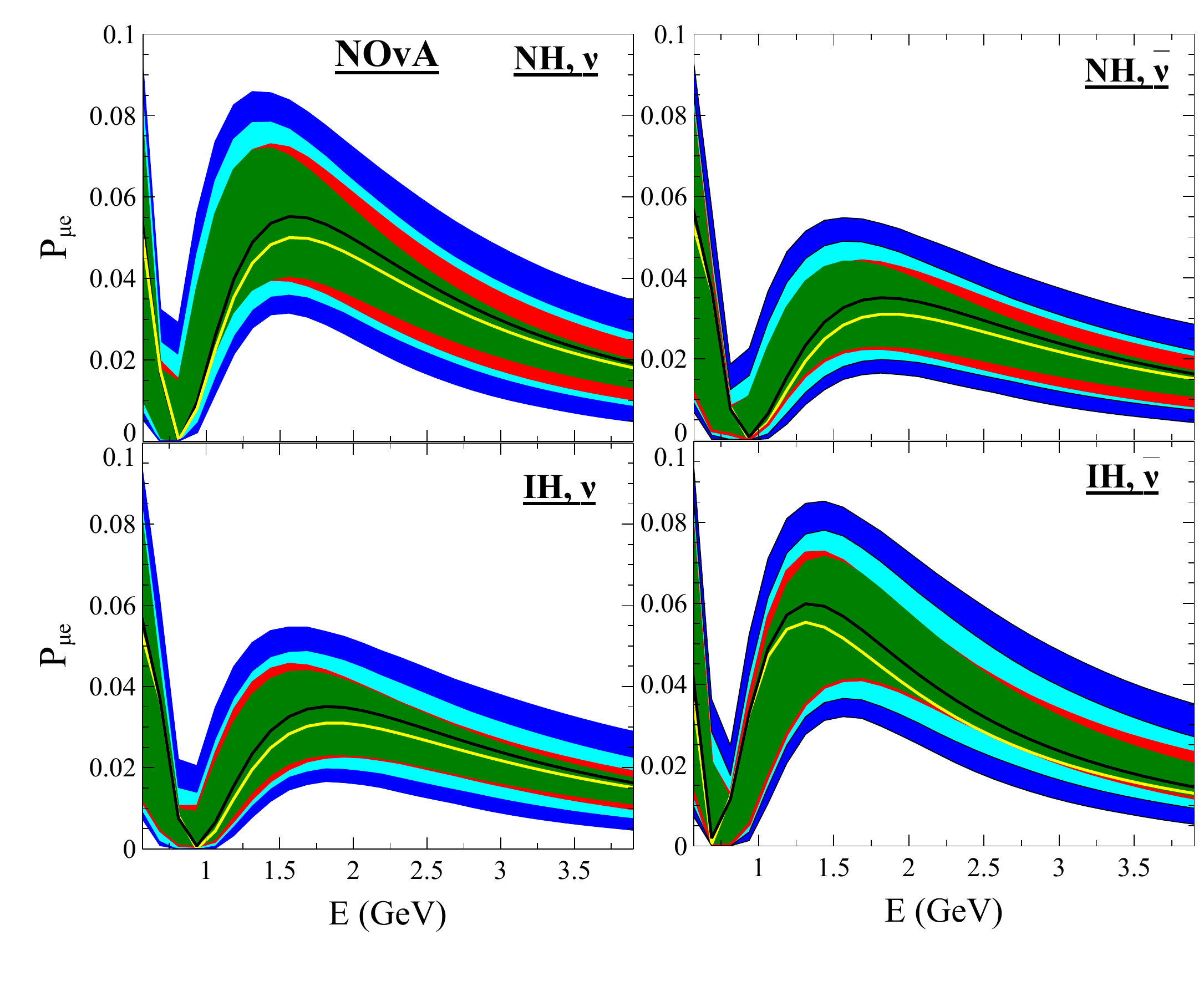}
\includegraphics[width=0.49\textwidth]{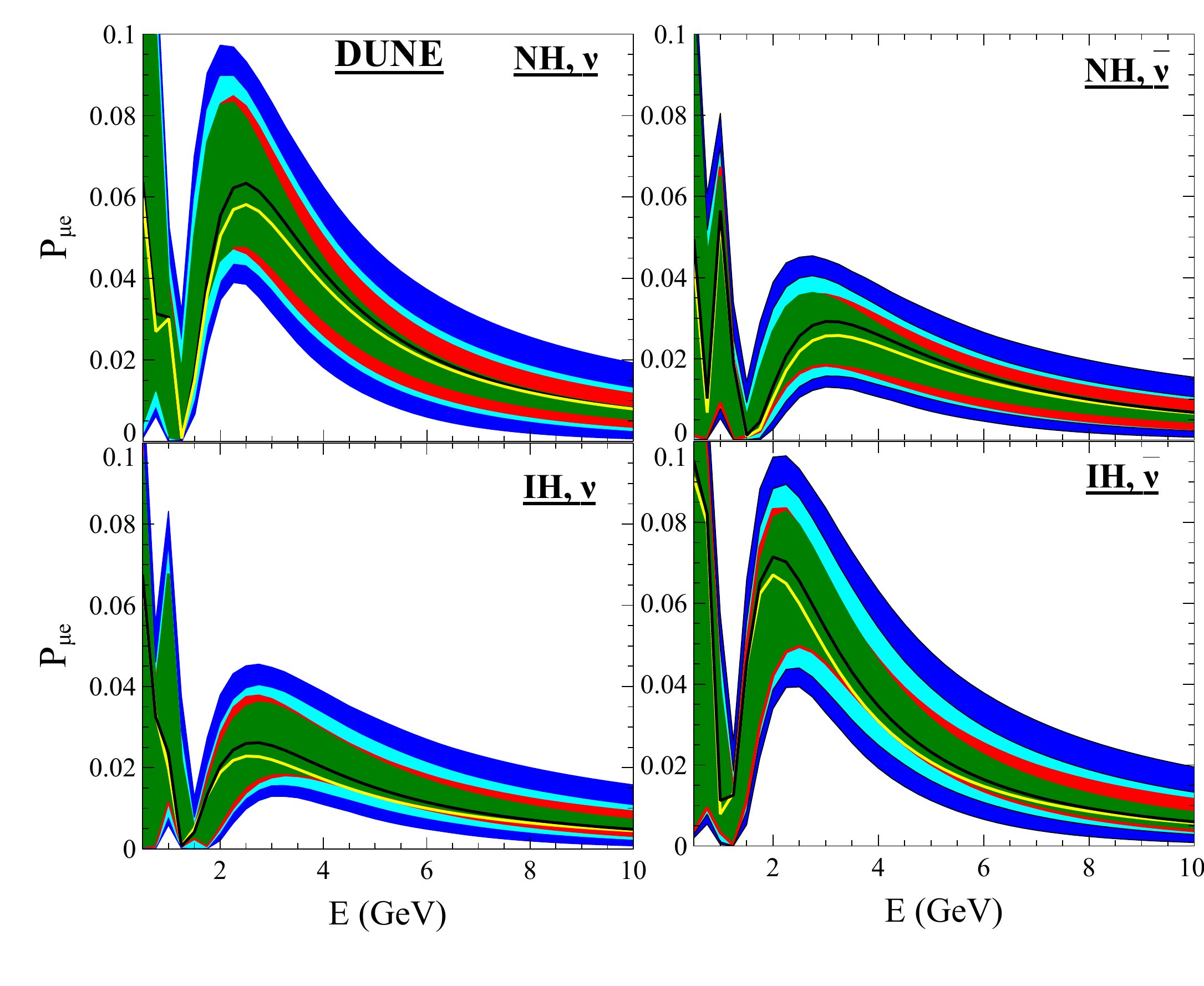}
\caption{\footnotesize{Variation of $\rm P(\nu_{\mu} (\bar{\nu_{\mu}})\rightarrow \nu_e (\bar{\nu_e}))$ with neutrino energy E for NO$\nu$A, T2K and DUNE: Plots are shown for two set of NU parameters $\alpha_{11}$, $\alpha_{22}$ and $|\alpha_{21}|$. The blue band (cyan and red band) corresponds to $\alpha_{11} = 0.9945$ (0.9973), $\alpha_{22} = 0.9995 $ (0.9998) and $|\alpha_{21}| = 0.0257$ (0.0128). The width of the blue and the cyan band is due to the variation of both $\delta_{CP}$ and the new phase $\phi_{NP}$ from $[-\pi, \pi]$ while that of the red and the green band is only because of $\delta_{CP}$ variation. The green band represents the standard three flavor oscillations scenario while $\delta_{CP}$ is varied in its full range $[-\pi, \pi]$. The black (w/o NU) and the yellow (with NU) lines correspond to $\delta_{cp} = 0^0$ and $\delta_{cp} = \phi_{NP} = 0^0$. }}
\label{prob}
\end{figure}

In FIG. \ref{prob}, the behaviour of the probability is depicted as a function of neutrino energy for the different experiment baselines. It is to be noted that the peak energy of the respective experiments is about 0.6 GeV for T2K, about 2 GeV for NO$\nu$A and about 3 GeV for DUNE. Both the neutrino and the antineutrino probabilities are depicted, under the assumption of normal neutrino mass hierarchy (NH) as well as inverted neutrino mass hierarchy (IH). The neutrino parameters are fixed at their respective best-fit values. The figures are plotted for 2 values of the non-unitary parameters - (a) with $\alpha_{11}$ and $\alpha_{22}$ at their respective lower bounds and $|\alpha_{21}|$ at its upper bound (blue band), and (b) with all three parameters at intermediate values within their allowed ranges (cyan band). The CP phase $\delta_{CP}$ and the non-unitary phase $\phi_{NP}$ are varied as described below.

In FIG. \ref{prob}, the width of a band for a given value of energy indicates the uncertainty in the oscillation probability due to the uncertainty in the parameter being varied. In the blue and cyan band, both $\delta_{CP}$ and the new phase $\phi_{NP}$ are varied over their full range, i.e. from $[-\pi, \pi]$. In the red and the green band, only $\delta_{CP}$ is varied over its full range $[-\pi, \pi]$. The green band represents the standard three-flavor (3+0) scenario, where all non-unitary parameters are set to zero, while in the red band the non-unitary parameters are non-zero (assumed to be at the intermediate values (b)), but the non-unitary phase $\phi_{NP}$ is zero. 

The following principal features can be observed from this figure:

\begin{itemize}

\item The deviation of the red band (non-zero NU parameters apart from NU phase) from the green band (standard 3-flavor case) is only because of the magnitude of the NU parameters, since the NU phase $\phi_{NP} = 0$ for the red band. It can be seen that the red and green bands have a minor difference over the energy range of interest, indicating that just the presence of non-unitarity might affect the determination of $\delta_{CP}$. The red band would be wider if more deviated values of NU parameters are taken.  

\item The difference between the red and green bands is nearly negligible for the Inverted Hierarchy case for all three baselines, while a small variation is visible in the Normal Hierarchy case. The difference gets accentuated in the energy range above the energy corresponding 
to the peak energy for each experiment. 

\item The blue and cyan bands are much wider compared to the red and green bands, and this indicates the increased uncertainty in the probability when the NU phase $\phi_{NP}$ is varied. As expected, the blue band is wider than the cyan band because the values of the NU parameters are at their bounds for the blue band and are less deviated from non-unitarity for the cyan band. The wide bands arising from variation of both $\delta_{CP}$ and $\phi_{NP}$ depicts the degeneracy between these two parameters, since for slightly displaced values of energy the same probability measurement may arise from different combinations of the two phases, effectively leading to a mimicking of the standard CP phase by the NU phase. Also, an observation of CP violation may be due to either of the two phases.

\item The black line corresponds to the CP invariant case in the 3+0 scenario ( $\delta_{cp} = 0^0$) while the yellow line is its counterpart in the new physics scenario (both $\delta_{cp} = \phi_{NP} = 0^0$). The NU parameters are kept at their central values for the latter case. It is observed that for T2K, both the plots are well separated while for NO$\nu$A and DUNE, the separation is distinct near the peak energy. This indicates that CP invariance in the $P_{\mu e}$ channel may be misinterpreted as CP violation in the three family scenario at these experiments.

\end{itemize}

 Further information may be derived from the CP (neutrino-antineutrino) asymmetry, which is defined as follows - 
$$
{\rm{A_{\nu{\bar{\nu}}} = (P_{\mu e} - \bar{P_{\mu e}}) / (P_{\mu e} + \bar{P_{\mu e}})}}
$$

 The CP asymmetry is a measure of CP violation, since it quantifies the change in the oscillation probability when the CP phase changes sign. In a CP conserving situation, ${\rm{A_{\nu{\bar{\nu}}}}}$ would be zero. The greater is the deviation of this factor from zero, the greater is the projected CP violation for that combination of parameters, baseline and energy. Also, the CP asymmetry takes into account the behaviour of the neutrino as well as antineutrino probabilities, and it has been noted that the combination of these two probabilities leads to special features which are not present in the neutrino or antineutrino probability alone \cite{26, 27}.

\begin{figure}[H]
\centering
\includegraphics[width=0.49\textwidth]{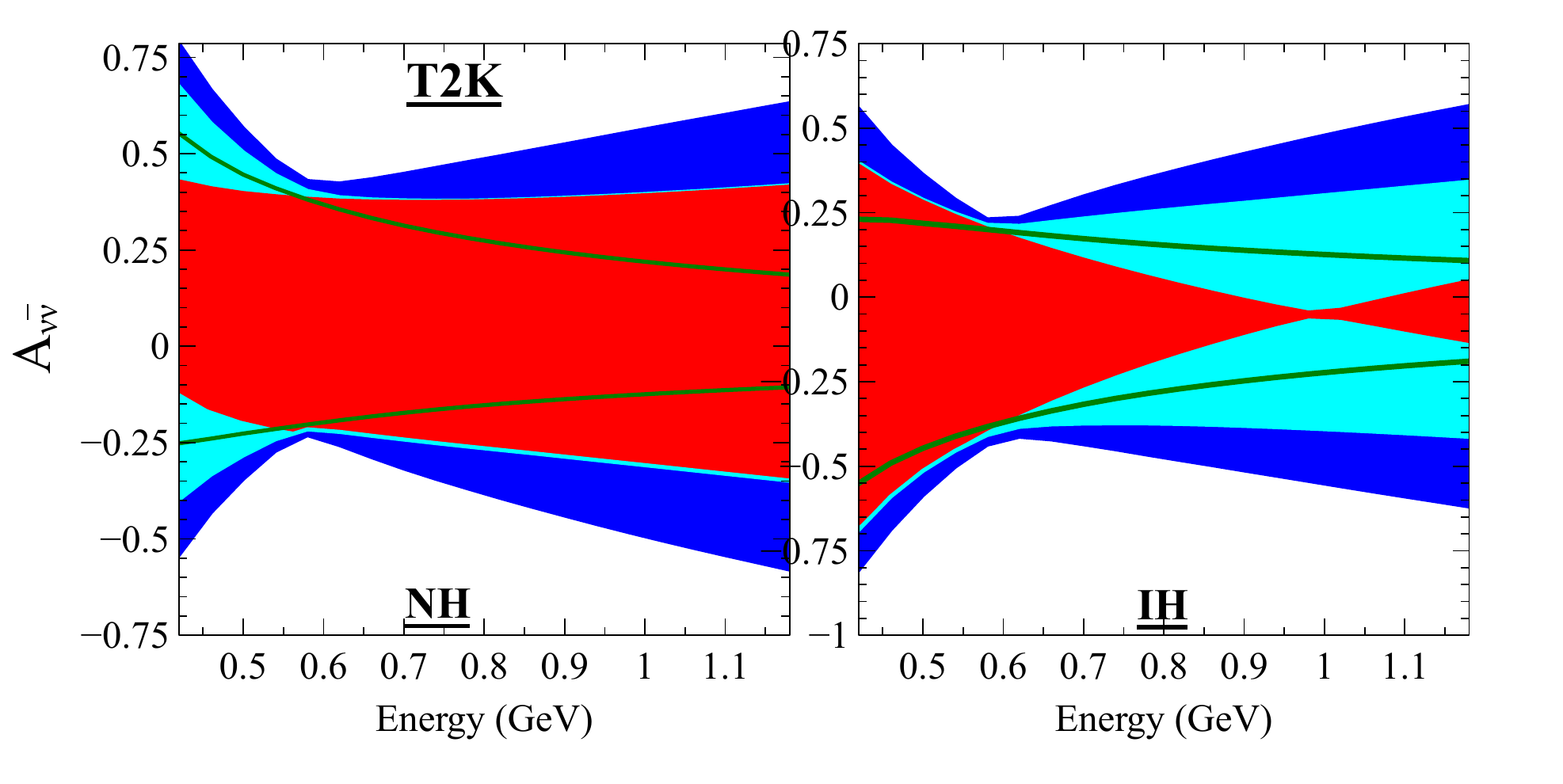}
\includegraphics[width=0.49\textwidth]{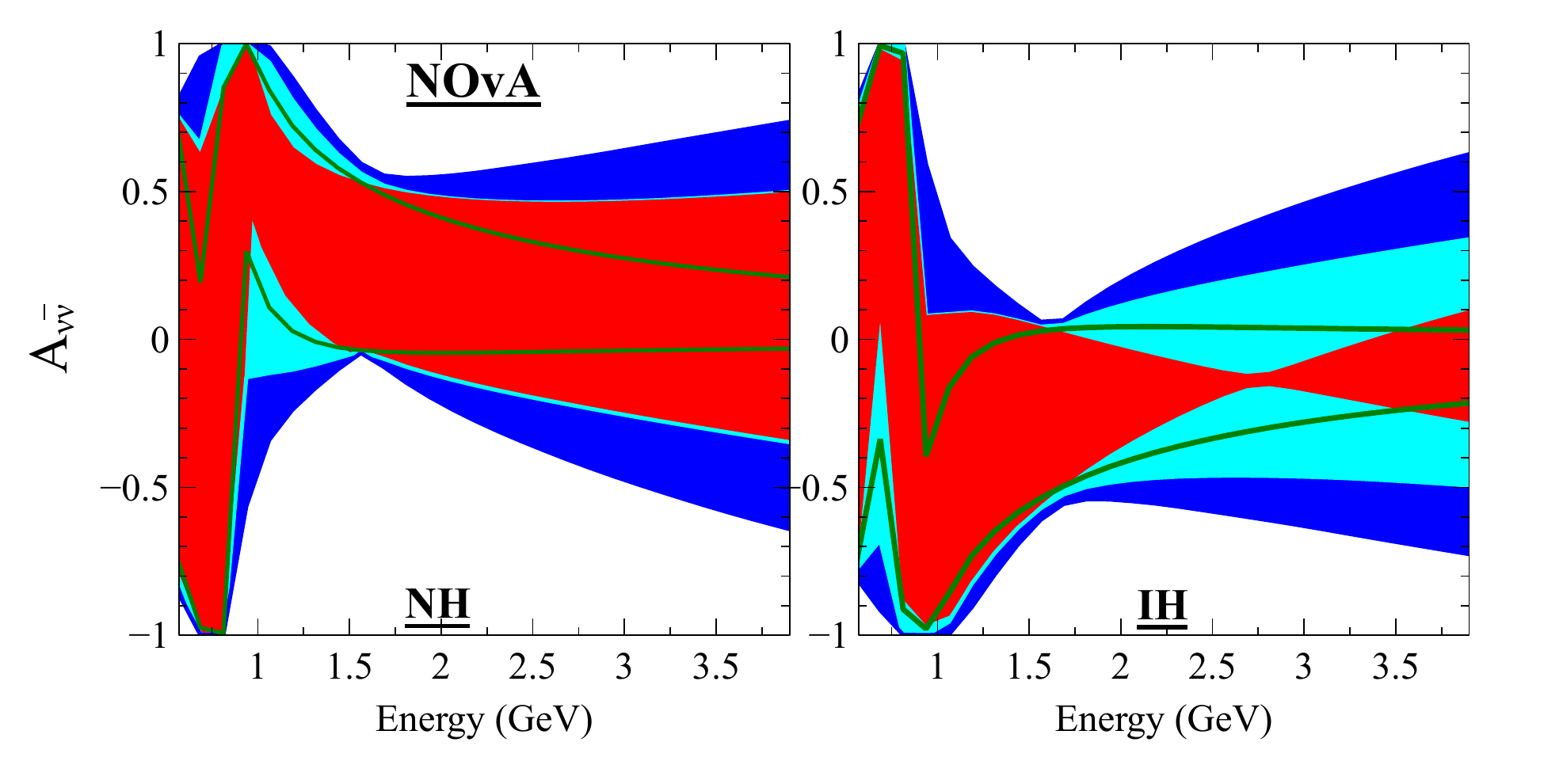}
\includegraphics[width=0.55\textwidth]{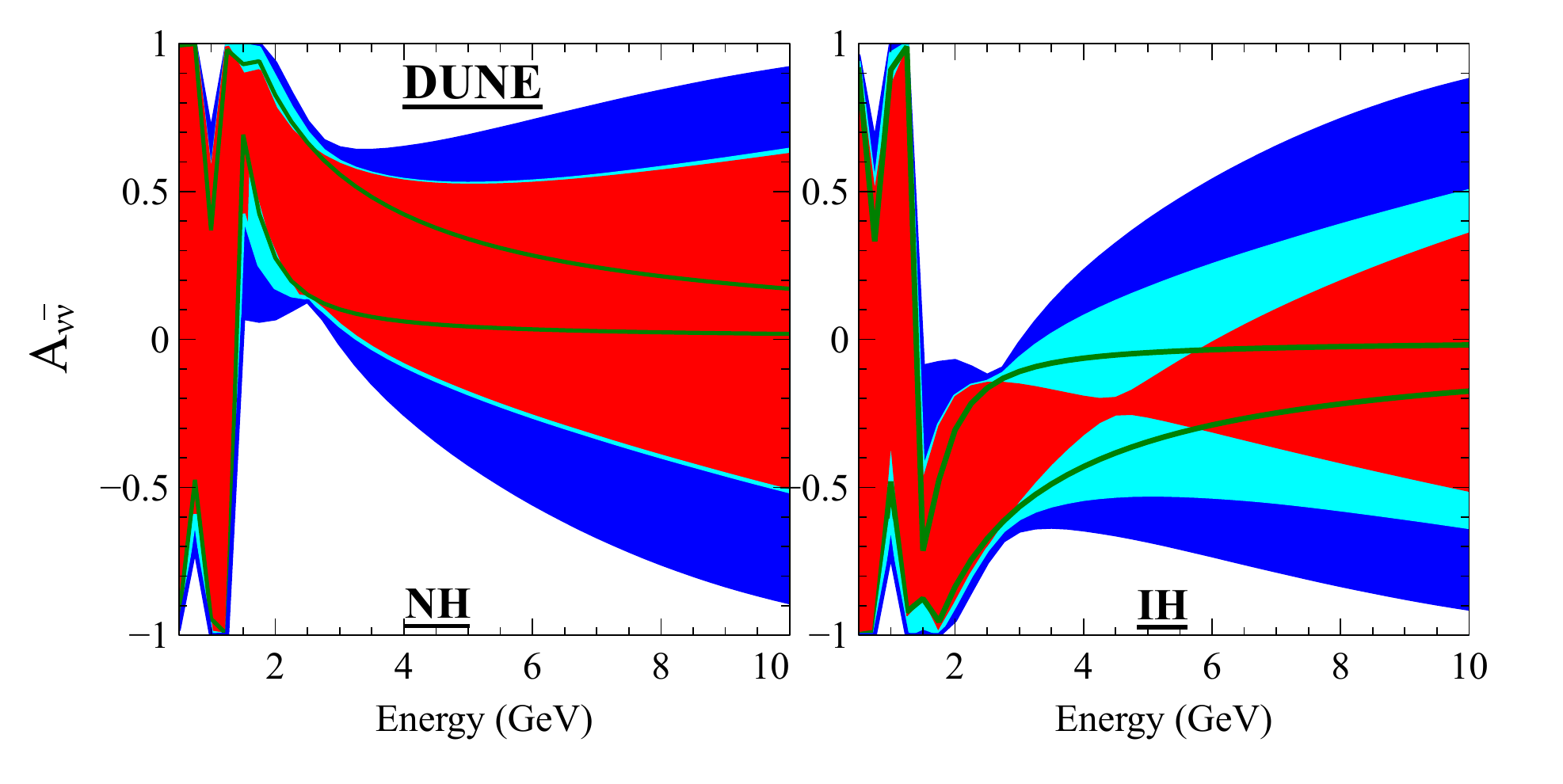}
\caption{\footnotesize{Variation of CP asymmetry $\rm A_{\nu\bar{\nu}}$ with Energy E for NO$\nu$A, T2K and DUNE: Plots are shown for two set of NU parameters $\alpha_{11}$, $\alpha_{22}$ and $|\alpha_{21}|$. The blue band (cyan and red band) corresponds to $\alpha_{11} = 0.9945$ (0.9973), $\alpha_{22} = 0.9995 $ (0.9998) and $|\alpha_{21}| = 0.0257$ (0.0128). The width of the blue and the cyan band is due to the variation of both $\delta_{CP}$ and the new phase $\phi_{NP}$ from $[-\pi, \pi]$ while that of the red is only because of $\delta_{CP}$ variation. The region between the two green lines represents the standard three flavor CP asymmetry.}}
\label{asym}
\end{figure}

 In FIG. \ref{asym}, the CP asymmetry is plotted as a function of neutrino energy for the three experiment baselines under consideration, for neutrinos and antineutrinos and for both normal and inverted mass hierarchy. As in FIG. \ref{prob}, the standard neutrino parameters are fixed at their best-fit values, and two sets of non-unitary parameters are chosen - putting the $\alpha$ parameters at their bounds (blue band) and at intermediate values (cyan band). In the blue and cyan bands, $\delta_{CP}$ as well as the NU phase $\phi_{NP}$ are varied over the full range [$-\pi,\pi$]. For the red band, $\delta_{CP}$ is varied over the full range and NU parameters are at intermediate values but $\phi_{NP}$ is set to zero. The region between the two green lines indicates the range of CP asymmetry for the 3-flavor scenario (all NU parameters zero).

The following prominent features may be noted from FIG. \ref{asym}:

\begin{itemize}

\item For the NH case, the region between the green lines (standard 3-flavor CP asymmetry) decreases gradually with increasing energy, while the width of the red band (CP asymmetry with non-zero NU parameters but $\phi_{NP} = 0$) increases with increasing energy. This means that at higher energies, the NU parameters, even without the NU phase, add to the uncertainty in the CP asymmetry. This behaviour is observed for all three baselines. 

\item At energy values of about 0.6 GeV for T2K, 1.6 GeV for NO$\nu$A and 2.6 GeV for DUNE, all 4 bands (red, green, blue and cyan) are seen to converge and become of the same width. This indicates that at such values of energy, the effect of NU parameters on the CP asymmetry is minimal and the variation due to NU parameters and $\phi_{NP}$ coincides with the asymmetry range due to $\delta_{CP}$ in the 3-flavour case. 
In \cite{26}, it was observed that a combination of neutrino and antineutrino channels helps to resolve the degeneracy between $\delta_{CP}$ and $\phi_{NP}$ for L/E $\sim$ 500 km/GeV, since some of the CP-dependent terms in the analytic probability expression cancel for this magnitude of L/E. It is interesting to note that the above energy values correspond to roughly L/E $=$ 500 km/GeV for the respective baselines. Hence this behaviour of the CP asymmetry reflects the cancellation of the $\delta_{CP}$-NU phase degeneracy in the CP asymmetry for these energies and baselines. Also, for the NH case, these energies act as a threshold after which the widths of the red band and the region between the green lines get reversed, with the red band becoming wider at higher energies. 

\item In all cases, the blue and cyan bands are narrowest at the energy values described above, and become wider as the energy is increased or decreased. The blue band is wider than the cyan band because the NU parameter values are at their bounds for the blue band and less deviated for the cyan band. The difference in width between the red and cyan bands is due to the NU phase $\phi_{NP}$, whose variation leads to a further uncertainty in the asymmetry. 

\item In the IH case, an interesting behaviour is observed. The red band, unlike for the NH case, becomes narrower with increasing energy at a faster rate than the region between the green lines (3-flavor asymmetry), and converges almost to zero at about 1 GeV for T2K, 2.7 GeV for NO$\nu$A and 4.5 GeV for DUNE. These energy values are somewhat removed from the peak energies of the respective experiments, and hence the behaviour in these regions may not have a major contribution to the CP violation sensitivities of the experiments. However, at the level of the oscillation probability this phenomenon is noteworthy, since it seems to indicate that the presence of non-unitarity leads to a nullification of the CP-violating behaviour of $\delta_{CP}$ for these specific energy and baseline combinations. In other words, the NU parameters, when $\phi_{NP}=0$, interact with the standard CP phase in a specific manner to cancel the effect of its variation on the CP asymmetry. The blue and cyan bands demonstrate their regular behaviour (increasing with energy) even at these energy values, which means that when $\phi_{NP}$ is varied the above effect no longer remains and the NU phase restores the uncertainty in the CP asymmetry which got removed due to the NU parameters. This behaviour is not easy to explain on the basis of the analytic probability expression, and we leave it for now as an intriguing observation which bears further study.    

\end{itemize}

In the next section we will analyze how these aspects of the oscillation probability behaviour with and without non-unitarity is reflected in the event rates and CP violation sensitivities of the given experiments. 

\section{Events and Sensitivity Studies}
\subsection{ Event Rate calculations}

In the previous section, we have discussed the effect of non-unitarity in CP violation measurements on the oscillation probability and the neutrino-antineutrino asymmetry. In this section we present a realistic analysis of the problem on the basis of event rate calculations. 
Here we have simulated all the three experiments separately using the details given in TABLE I and TABLE II.

\begin{figure}[t]
\centering
\includegraphics[width=0.49\textwidth]{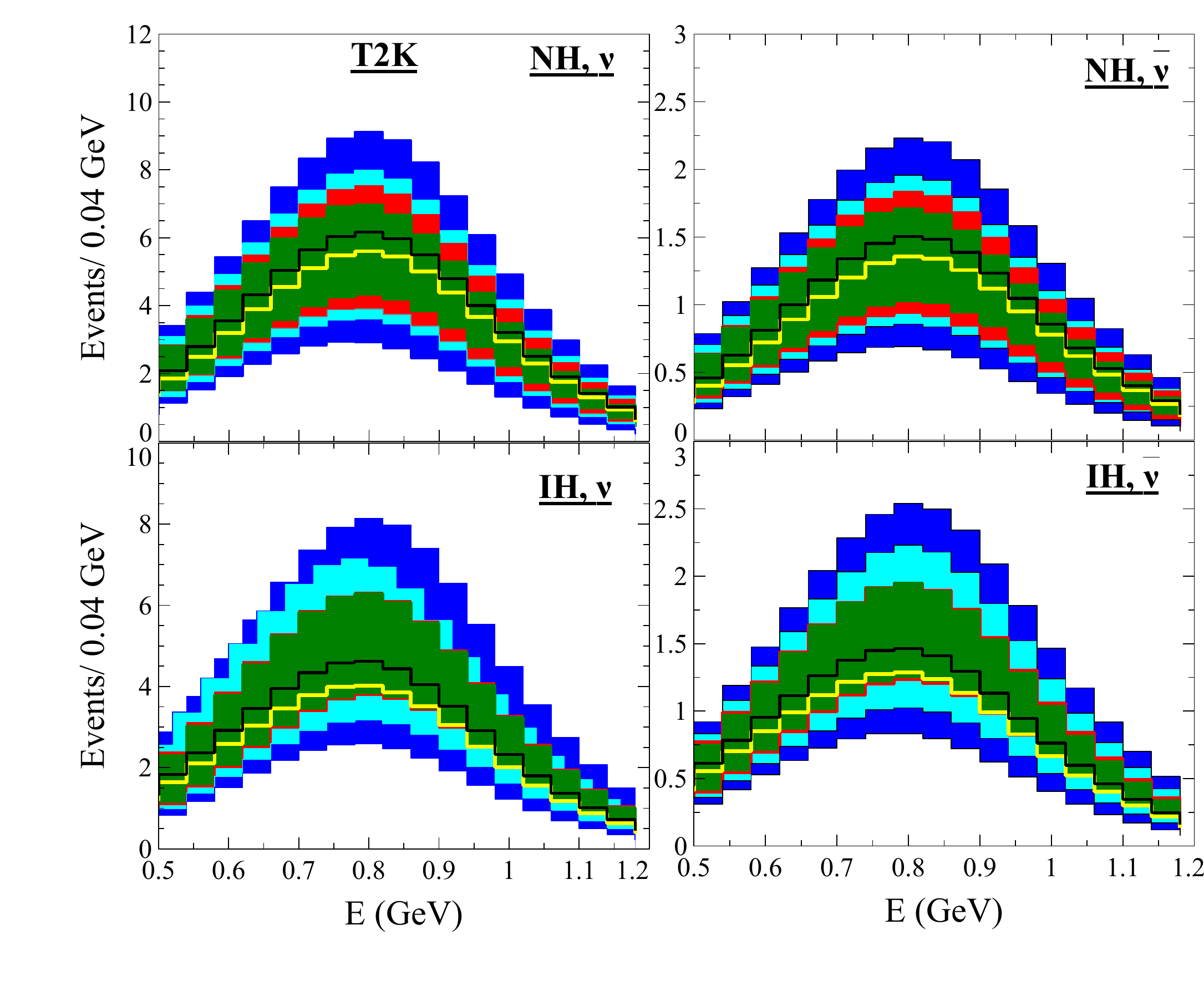}
\includegraphics[width=0.49\textwidth]{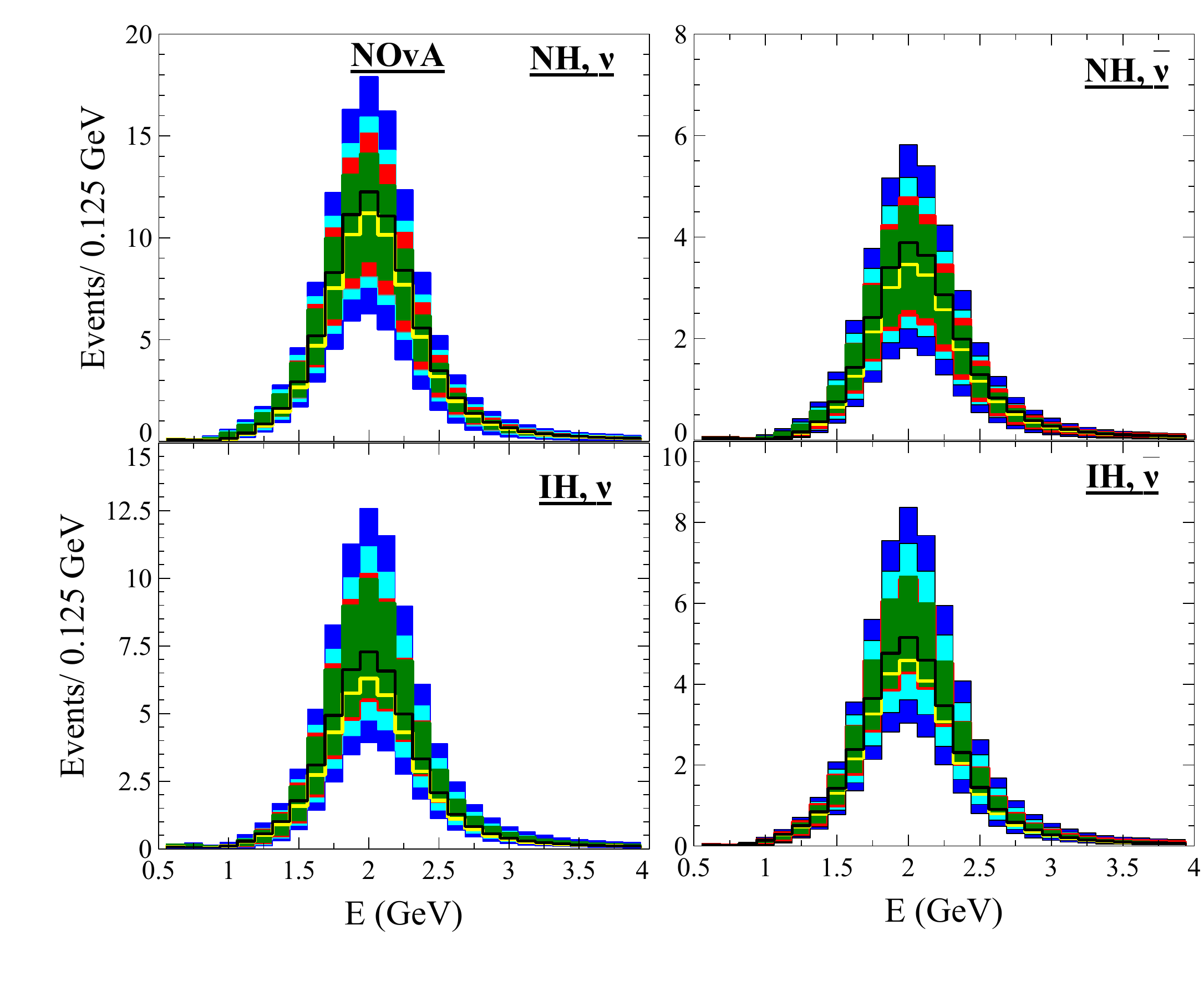}
\includegraphics[width=0.49\textwidth]{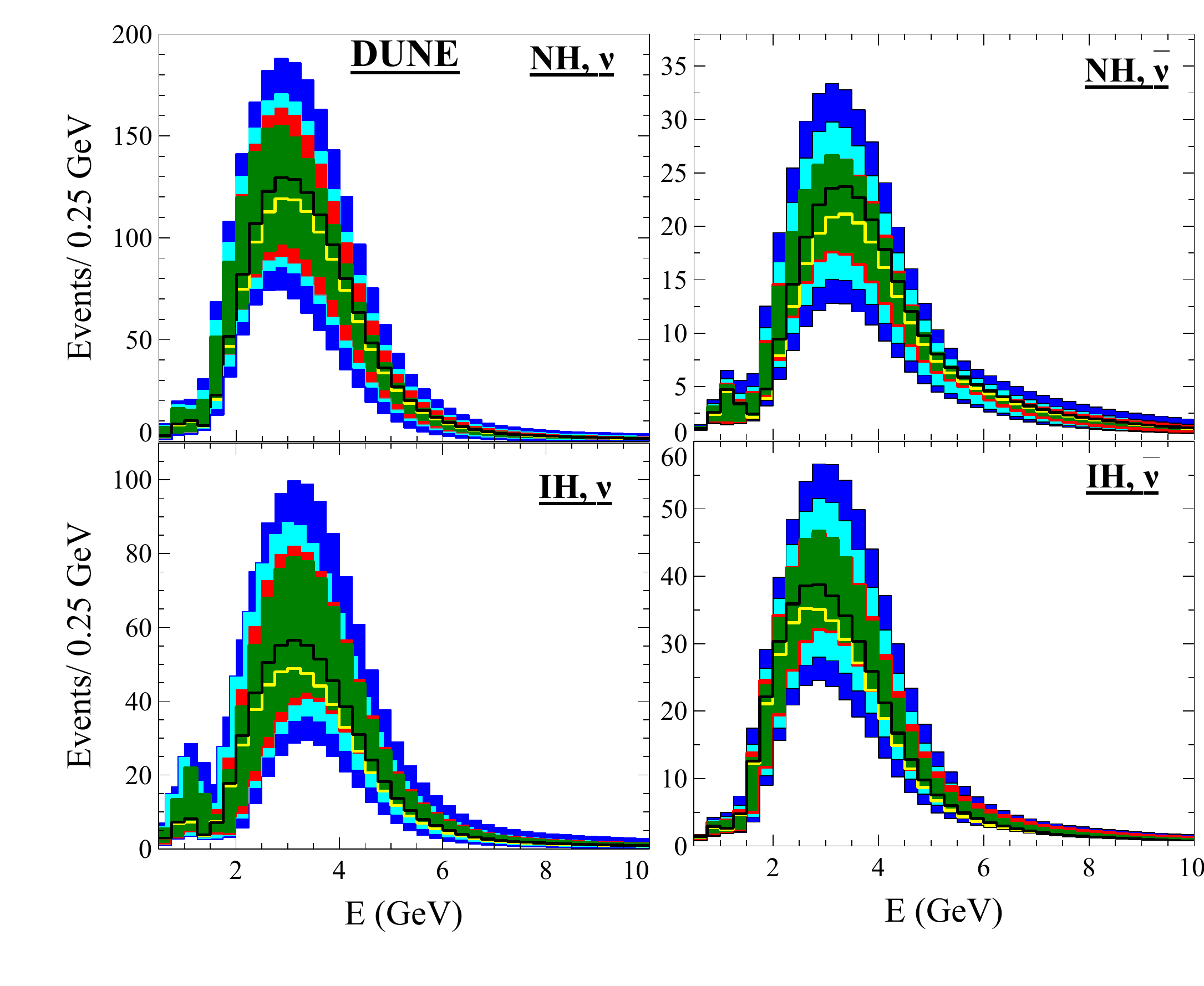}
\caption{\footnotesize{Variation of binned event rates with reconstructed neutrino energy E for NO$\nu$A, T2K and DUNE for two sets of NU parameters both in neutrino and antineutrino mode for both the hierarchies. The blue band (cyan and red band) corresponds to $\alpha_{11} = 0.9945$ (0.9973), $\alpha_{22} = 0.9995 $ (0.9998) and $|\alpha_{21}| = 0.0257$ (0.0128). The green band is the standard 3+0 band.  The black (w/o NU) and the yellow (with NU) lines correspond to $\delta_{cp} = 0^0$ and $\delta_{cp} = \phi_{NP} = 0^0$. }}\label{CPdune}
\end{figure}

 We plot the binned events as a function of reconstructed energy. The green band is the standard 3+0 band which arises due to the variation of $\delta_{CP}$ from $-\pi$ to $\pi$. The blue and the cyan bands arise due to the effect of non-unitarity, when we vary both $\delta_{CP}$ and the new phase $\phi_{NP}$ in the range $[-\pi, \pi]$. The red band is a subset of the cyan band obtained by putting $\phi_{NP} = 0$. The behaviour of the events for all the experiments is consistent with the probability behaviour. We can draw the following conclusions from the event plots: 
 \begin{itemize}
 \item For any set of values of the NU parameters, the 3+0 band is always bounded by the bands with unitarity violation, and hence in the overlapping region i.e. in the green region, it is not possible to ascertain whether the events are due to variations of NU parameters or of $\delta_{CP}$ in the 3+0 scenario.
 
 \item The degeneracy still persists even when the NU phase is 0 i.e. $\phi_{NP} = 0$. Not only the new phase $\phi_{NP}$, but all the three other NU parameters $\alpha_{11}$, $\alpha_{22}$ and $|\alpha_{21}|$, also play a role in maintaining the ambiguity. The more we deviate from unitarity, the more is the width of the red band, indicating substantial degeneracy. 
 
 \item The behaviour of the event plots in terms of the band placements is similar for the different experiments, for both the hierarchies and both in neutrino and antineutrino mode.
 
 \item The behaviour of the event plots for the CP invariant case, i.e.  $\delta_{cp} = 0^0$ (black) and $\delta_{cp} = \phi_{NP} = 0^0$ (yellow), is consistent with the corresponding probability plots. It is observed that both the lines lie within the green band for all the three experiments and hence CP invariance may be confused with CP violation in the case of three family neutrino oscillations.
 \end{itemize}
 Thus in the presence of non-unitarity, the measurement of $\delta_{CP}$ and hence of CP violation gets affected. In the succeeding section, we will perform a $\chi^2$ analysis to discuss the effects of this ambiguity in detail 
in the light of the different experiments and their combination.
\subsection{Statistical Details and $\chi^2$ Analysis}

We define the $\chi^2$ (statistical) for CP violation sensitivity as follows: 
\begin{equation}
\chi^2 = \sum_{i=1}^{bins} \sum_{j}^{2}\frac{[N^{i,j}_{true} - N^{i,j}_{test}]^2}{N^{i,j}_{true}}
\end{equation}
, where $N^{i,j}_{true}$ and $N^{i,j}_{test}$ are the event rates that correspond to data and fit in the $i^{th}$ bin. $j = 1$ is for neutrinos and $j = 2$ for anti-neutrinos. The number of bins are different for each experiment i.e. for DUNE there are 39 bins each of width 250 MeV in the energy range 0.5 to 10 GeV, for NO$\nu$A there are 28 bins of width 125 MeV in the energy range 0.5 to 4 GeV and for T2K, we have 20 bins of width 40 Mev in the range 0.4 to 1.2 GeV. 

 In this work our main focus is to point out the effect of the new phase on CP violation measurements at DUNE, NO$\nu$A and T2K, and hence to study the CP violation sensitivity as well as CP discovery potential of these experiments in the presence of this new phase. The capability of an experiment to differentiate between CP conserving and CP violating values of $\delta_{CP}$ is a measure of its CP sensitivity. So in the standard three family case (SI), $N^{i,j}_{test}$ is only dependent on $\delta_{CP}$ as we have not marginalised over any other three flavor oscillation parameters throughout this work (assuming them to be known precisely). Hence, while calculating the CP sensitivity we fix $\delta_{CP}$ at 0 and $\pi$ in the `fit' and vary over the whole range of $\delta_{CP}$ from -$\pi$ to $\pi$ in `data' assuming normal hierarchy as the true hierarchy. When unitarity is violated, we have an additional phase and three new additional parameters $\alpha_{11}$, $\alpha_{22}$ and $|\alpha_{21}|$. Hence in the `fit', we take all possible combinations of 0 and $\pi$ (4 combinations) for both $\delta_{CP}$ and the new phase $\phi_{NP}$ and vary $\alpha_{11}$, $\alpha_{22}$ and $|\alpha_{21}|$ in their allowed range. We then calculate the minimised $\chi^2$ i.e $\chi^2_{\rm min}$ as a function of the true parameters and choose the minimum and maximum of $\chi^2_{\rm min}$ for a given true value of $\delta_{CP}$. We have also marginalised over the systematic uncertainties for each experiment as given in TABLE I.
 
  To analyze the effect of the non-unitarity phase on the measurement of CP violation sensitivity, we have also calculated the standard CP violation sensitivity marginalising over the whole range of $\phi_{NP}$ in the `fit' along with all the NU parameters.


\subsection{ CP Violation Sensitivity}
While studying the CP violation sensitivity, it is important to note that the $P_{\mu e}$ oscillation probability is principally associated with CP violation since it is a function of the CP odd term $\sin \delta_{CP}$. But due to the presence of the CP even term $\cos\delta_{CP}$, $P_{\mu\mu}$ can contribute slightly to CP sensitivity through statistics. As we are studying the potential of each experiment to determine CP violation sensitivity, we have considered both appearance and disappearance channels in the simulation. In this section, we discuss two scenarios related to CP violation measurements. 
\begin{enumerate}
\item CP violation irrespective of the source i.e. originating either from the standard Dirac $\delta_{CP}$ phase or from the non-unitary phase $\phi_{NP}$.
\item 'Standard' CP violation due to the standard Dirac $\delta_{CP}$ phase only, in the presence of non-unitarity.
\end{enumerate}
To analyze Case (1), as discussed in the preceding section, we consider four possible combinations of $\delta_{CP}$ and $\phi_{NP}$ in `fit', each of which can be either 0 or $\pi$. 
\begin{figure}[t]
\centering
\includegraphics[width=1.0\textwidth]{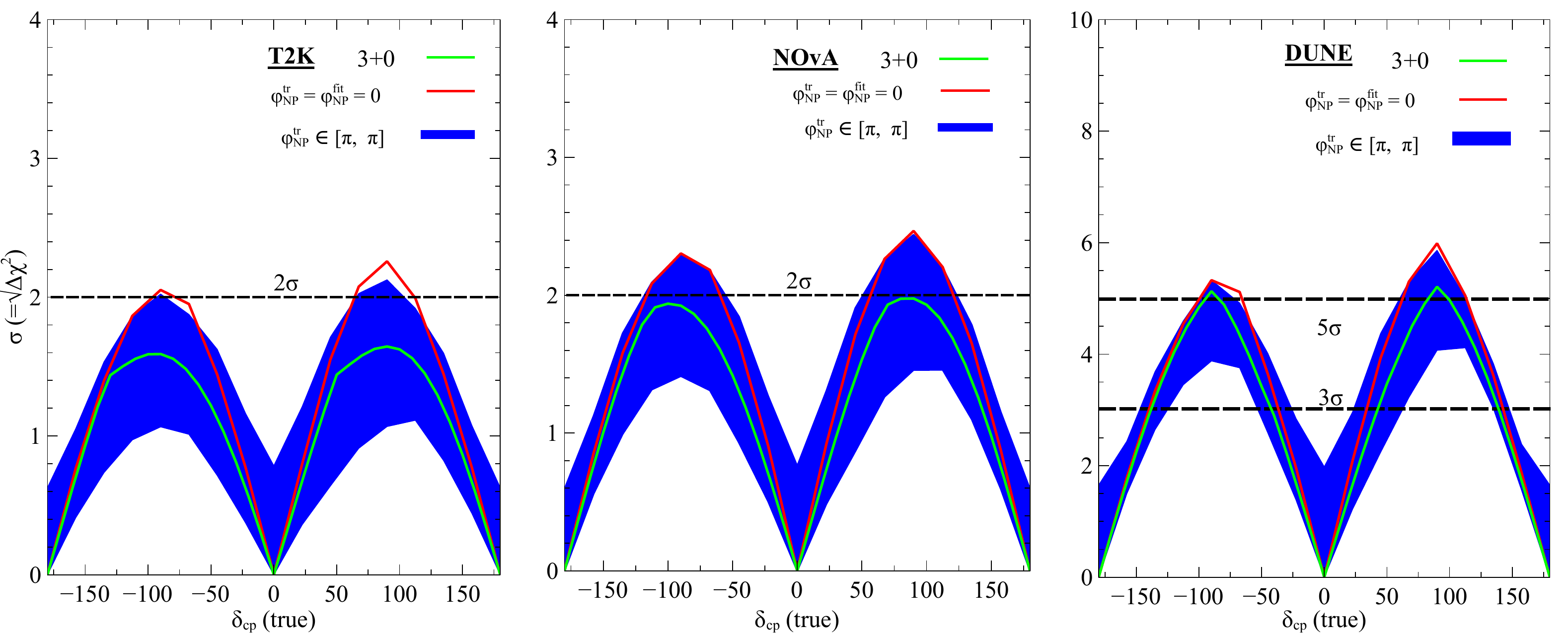}
\caption{\footnotesize{CP violation sensitivity for T2K, NO$\nu$A and DUNE as a function of true $\delta_{CP}$. The width of the blue band is because of variation of true $\phi_{NP}$. The red plot is a special case of the blue band when $\phi_{NP}=0$ both in `data' and `fit'. The green plot represents the 3+0 CP violation sensitivity. }}
\label{cp1}
\end{figure}
In FIG. \ref{cp1} We show the CP violation sensitivity of the three superbeam experiments DUNE, NO$\nu$A and T2K individually. For comparison, we have also included their 3+0 CP violation results. It is observed for all the three experiments that in the presence of non-unitarity, the ability to measure the CP violation sensitivity changes drastically. We have assumed the central values of the allowed range of the non-unitarity parameters, $\alpha_{11}$, $\alpha_{22}$ and $|\alpha_{21}|$ as the true values and for this particular choice of `data', CP violation sensitivity crosses to both sides of the 3+0 plot. This behavior can be explained on the basis of two effects: 
\begin{itemize}
\item Effect of large parameter space
\item Effect of deviation from unitarity
\end{itemize}
In the 3+0 case, we have only one $\delta_{CP}$ phase and hence to study CP violation sensitivity we contrast any true $\delta_{CP}$ with CP conserving phases 0 and $\pi$ in `fit'. But in the presence of non-unitarity, we have a total of 4 new parameters and hence 5 parameters to be marginalised in the `fit' i.e.  $\alpha_{11}$, $\alpha_{22}$, $|\alpha_{21}|$, $\delta_{CP}$ and $\phi_{NP}$. So marginalising over a large parameter space reduces the $\chi^2_{\rm min}$ value. But at the same time, as we deviate from unitarity (i.e. $\alpha_{11} =1.0$, $\alpha_{22} = 1.0$, $|\alpha_{21}| = 0$), the band corresponding to $\chi^2_{\rm min}$ starts to broaden. Specifically, with an increase in $|\alpha_{21}|$, the variation in the true values of  $\phi_{NP}$ also increases. The band width increases as the deviation increases and it spreads on both sides of the 3+0 curve. In the case depicted, the second effect is dominating over the first and hence the sensitivity to CP violation is higher than that of 3+0 for some combinations of $(\delta_{CP}, \phi_{NP})$. As a result, even at the CP conserving phases 0 and $\pm \pi$, there is some CP violation sensitivity. This arises from the fact that the true value of $\phi_{NP}$ is not necessarily CP conserving at these points. In DUNE, it is almost a $2\sigma$ effect. All the three experiments show a uniform response. The red curve shows the sensitivity when the new phase variation is absent i.e. $\phi_{NP}= 0$ and the true values of the NU parameters are still at the central values of their allowed ranges. 
\begin{figure}[t]
\centering
\includegraphics[width=0.49\textwidth]{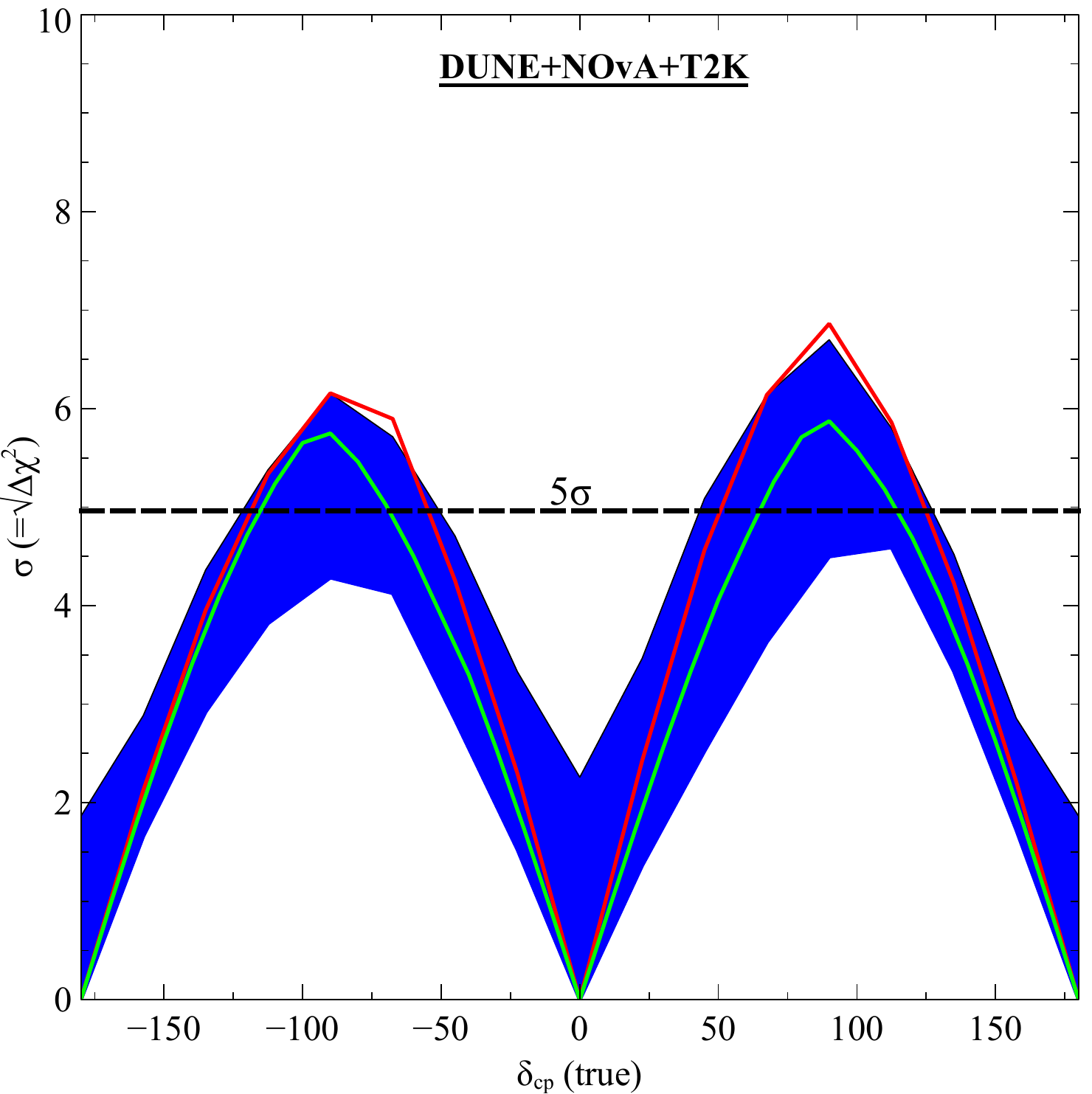}
\caption{\footnotesize{CP violation sensitivity for the combined set up as a function of true $\delta_{CP}$. The width of the blue band is because of variation of true $\phi_{NP}$. The red plot is a special case of the blue band when $\phi_{NP}=0$ both in `data' and `fit'. The green plot represents the 3+0 CP violation sensitivity.}}
\label{cpCom}
\end{figure}
In FIG. \ref{cpCom}, we show the sensitivity to CP violation when the three experiments are combined. As expected, combining the experiments enhances the sensitivity, but the degeneracy described above is still present. 

\begin{figure}[H]
\centering
\includegraphics[width=1.0\textwidth]{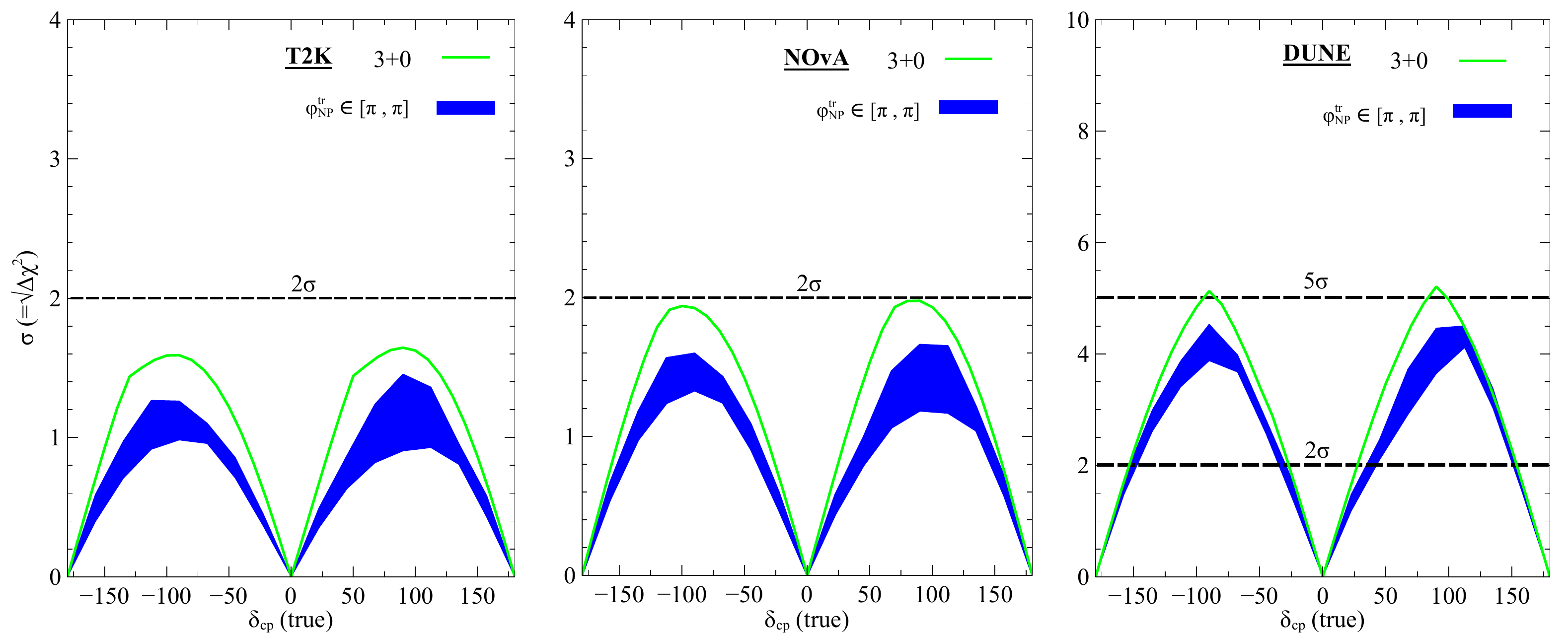}
\caption{\footnotesize{ $\delta_{CP}$ originated sensitivity plots for T2K, NO$\nu$A and DUNE as a function of true $\delta_{CP}$ in presence of non-unitarity. The width of the blue band is because of variation of true $\phi_{NP}$. The green plot represents the 3+0 CP violation sensitivity. }}
\label{cp_margi}
\end{figure}

\begin{figure}[H]
\centering
\includegraphics[width=0.49\textwidth]{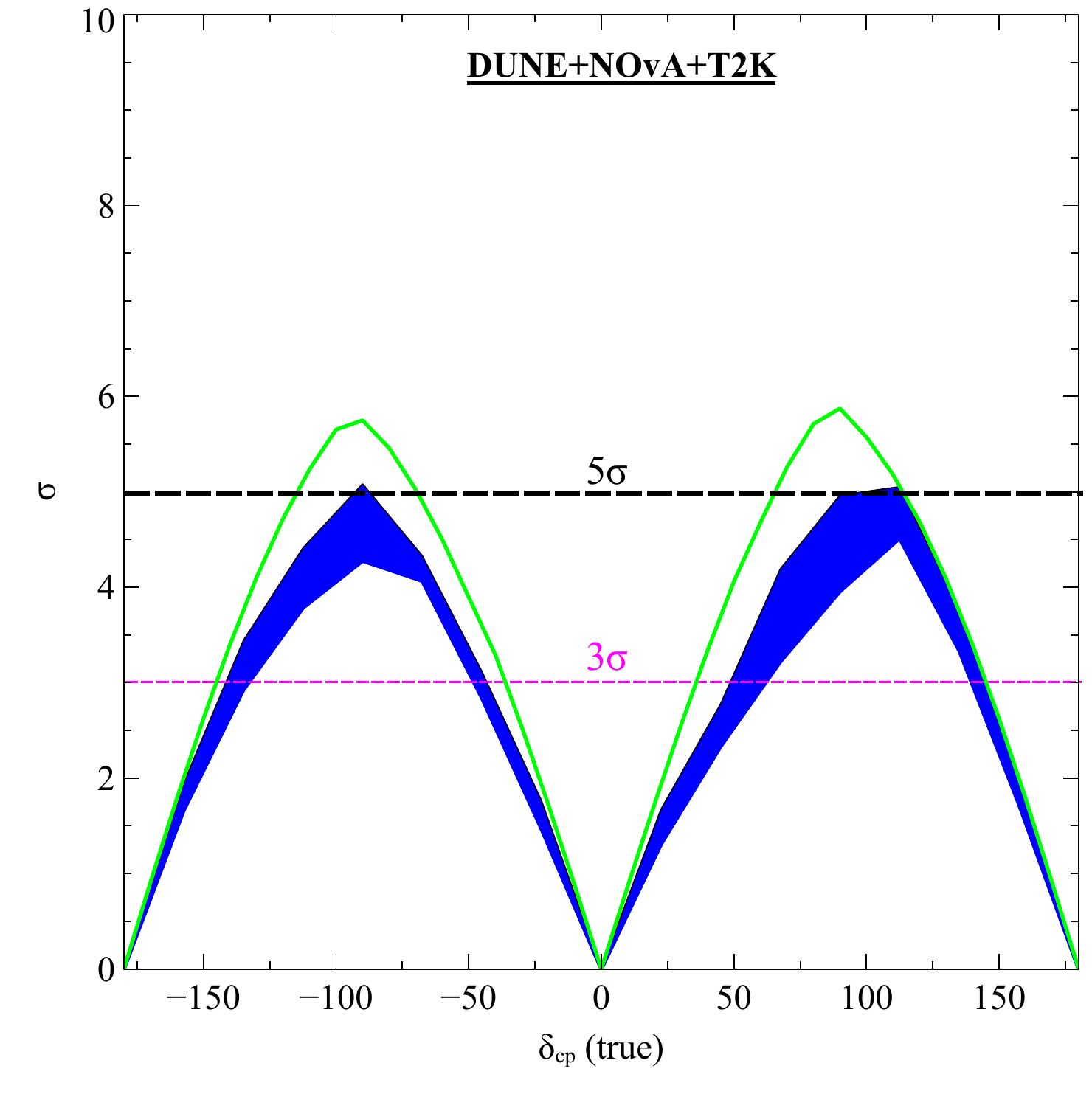}
\caption{\footnotesize{$\delta_{CP}$ originated sensitivity plots for for the combined set up as a function of true $\delta_{CP}$ in presence of non-unitarity. The width of the blue band is because of variation of true $\phi_{NP}$. The green plot represents the 3+0 CP violation sensitivity.}}
\label{cpCom2}
\end{figure}

Alternatively, we may formulate the problem in a slightly different way by asking the question differently. In the presence of non-unitarity, how sensitive are these experiments  to 'standard' CP violation that originates only from the standard Dirac $\delta_{CP}$ phase? This corresponds to Case (2) listed above. To answer this question, we choose only 0 and $\pi$ in `fit' for $\delta_{CP}$ and vary $\phi_{NP}$ in the full range both in `data' and `fit'. All other NU parameters are marginalised as before. FIG. \ref{cp_margi} and FIG. \ref{cpCom2} show the individual as well as the combined results for all the three experiments under this assumption.
The following conclusions can be drawn from this figure:
\begin{itemize}
\item The presence of non-unitarity hampers the standard CP violation sensitivity and it decreases.
\item There is a distinct separation between the 3+0 CP sensitivity plot and the band that corresponds to standard CP violation sensitivity in the presence of non-unitarity for most of the true values of $\delta_{CP}$ 
\item Combining all the three experiments can enhance the sensitivity but still the sensitivity is less compared to that of the 3+0 combined sensitivity. 
\item The effect of large parameter space dominates the effect of deviation from unitarity. Due to this, $\chi^2_{\rm min}$ and hence the CP violation sensitivity decreases uniformly compared to the 3+0 curve.  
\end{itemize}
\begin{figure}[t]
\centering
\includegraphics[width=0.35\textwidth]{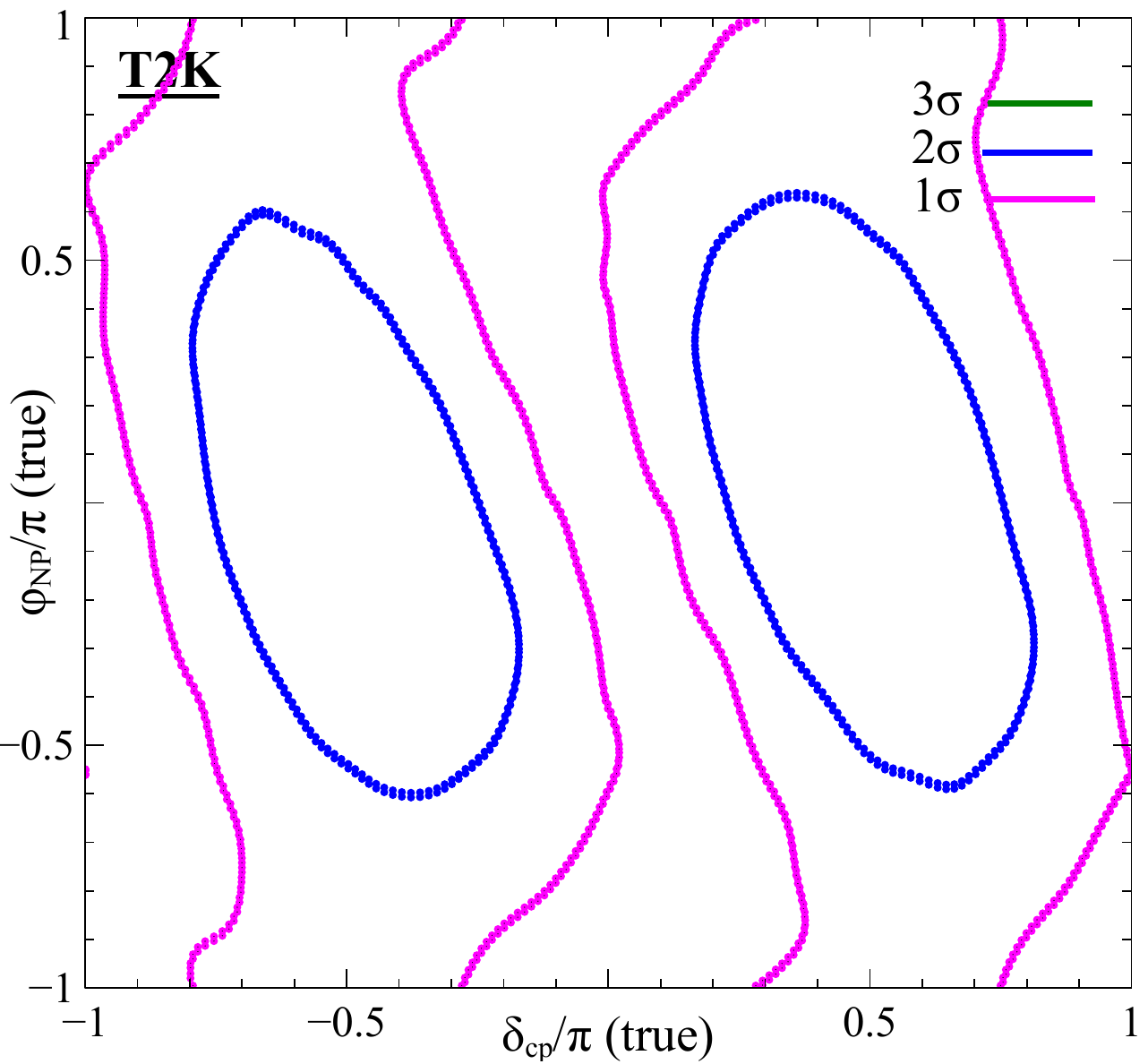}
\includegraphics[width=0.35\textwidth]{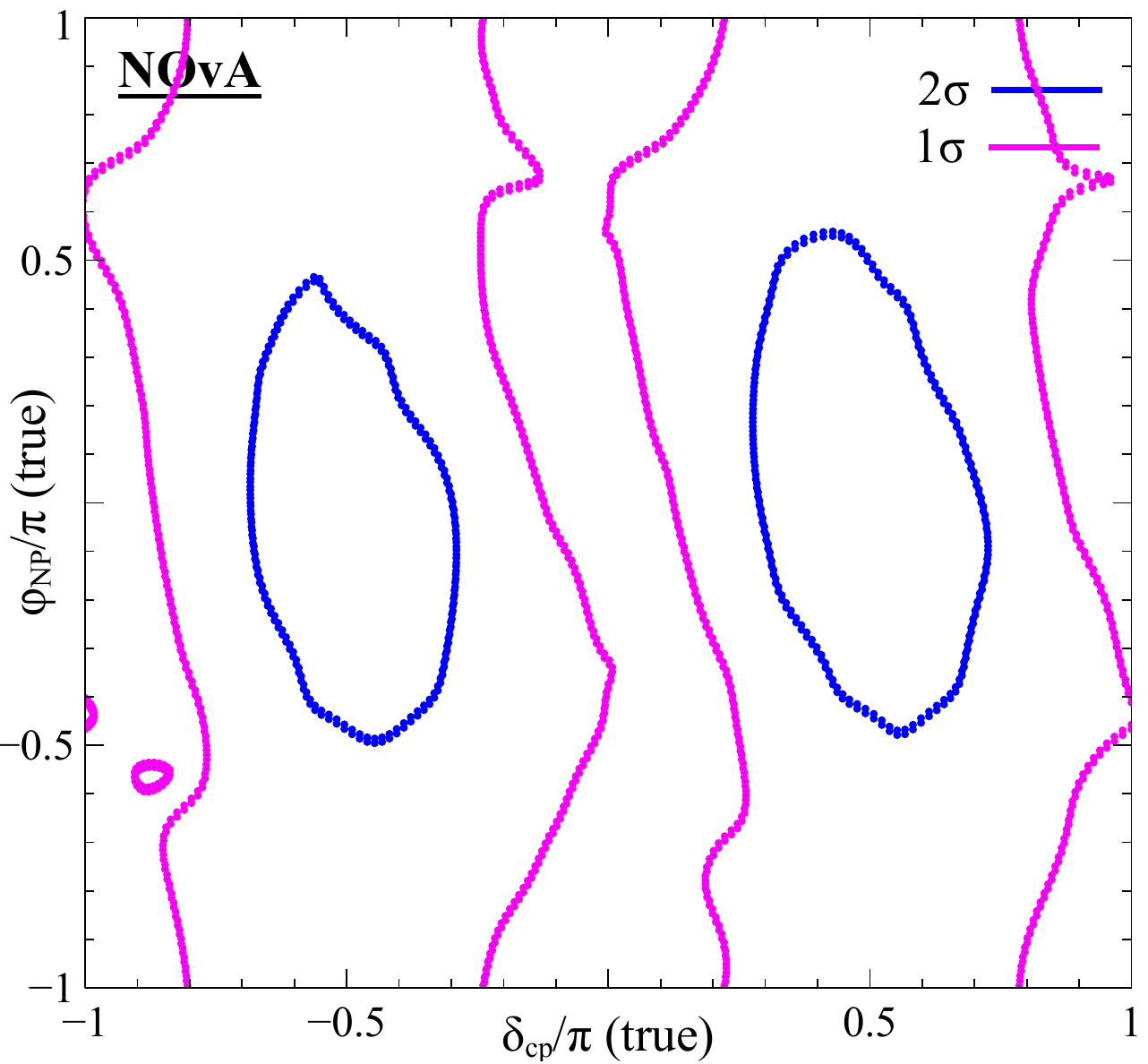}
\includegraphics[width=0.35\textwidth]{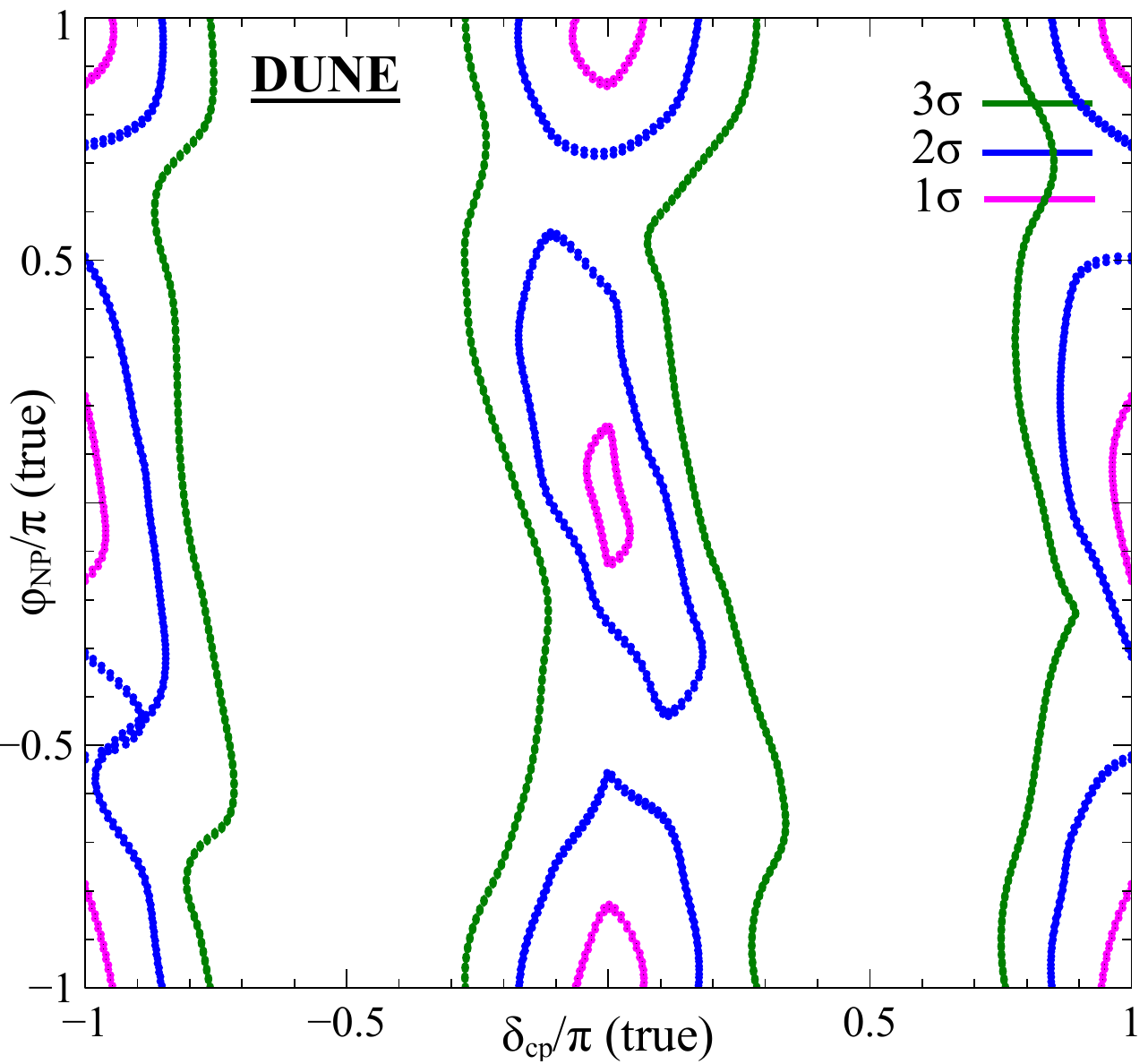}
\includegraphics[width=0.35\textwidth]{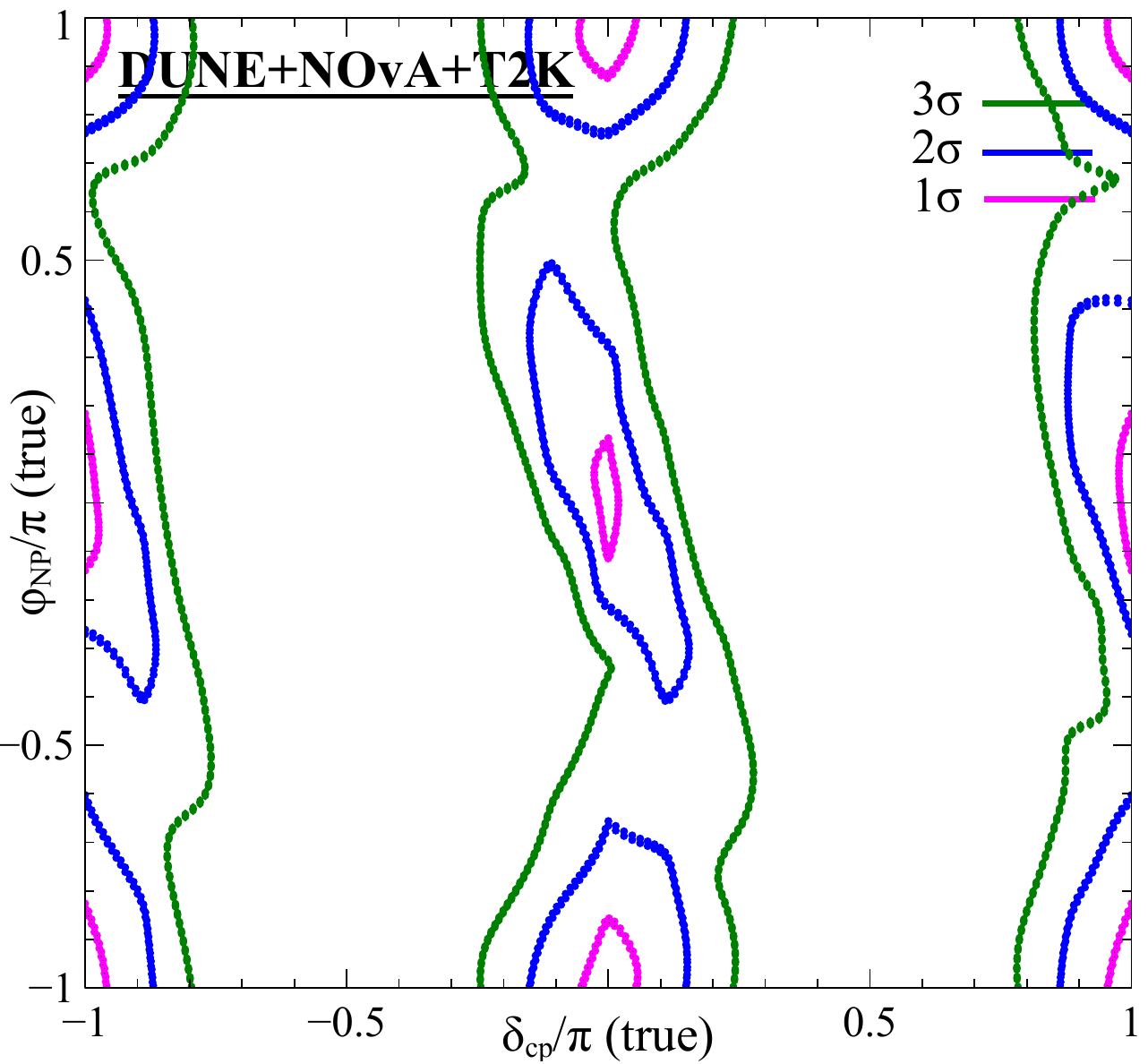}
\caption{\footnotesize{CPV discovery plots for T2K, NOvA, DUNE and their combined set ups, irrespective of  the source of CP. Here in the fit, we have assumed the CP conserving values 0 and $\pi$ for both $\delta_{CP}$ and $\phi_{NP}$ and hence for each set of true values of $\delta_{CP}$ and $\phi_{NP}$, we have total 4 choices of combinations of  0 and $\pi$ in the fit. We have also marginalized over all the non-unitarity parameters in their allowed range.}}
\label{cpdis1}
\end{figure}
\begin{figure}[t]
\centering
\includegraphics[width=0.35\textwidth]{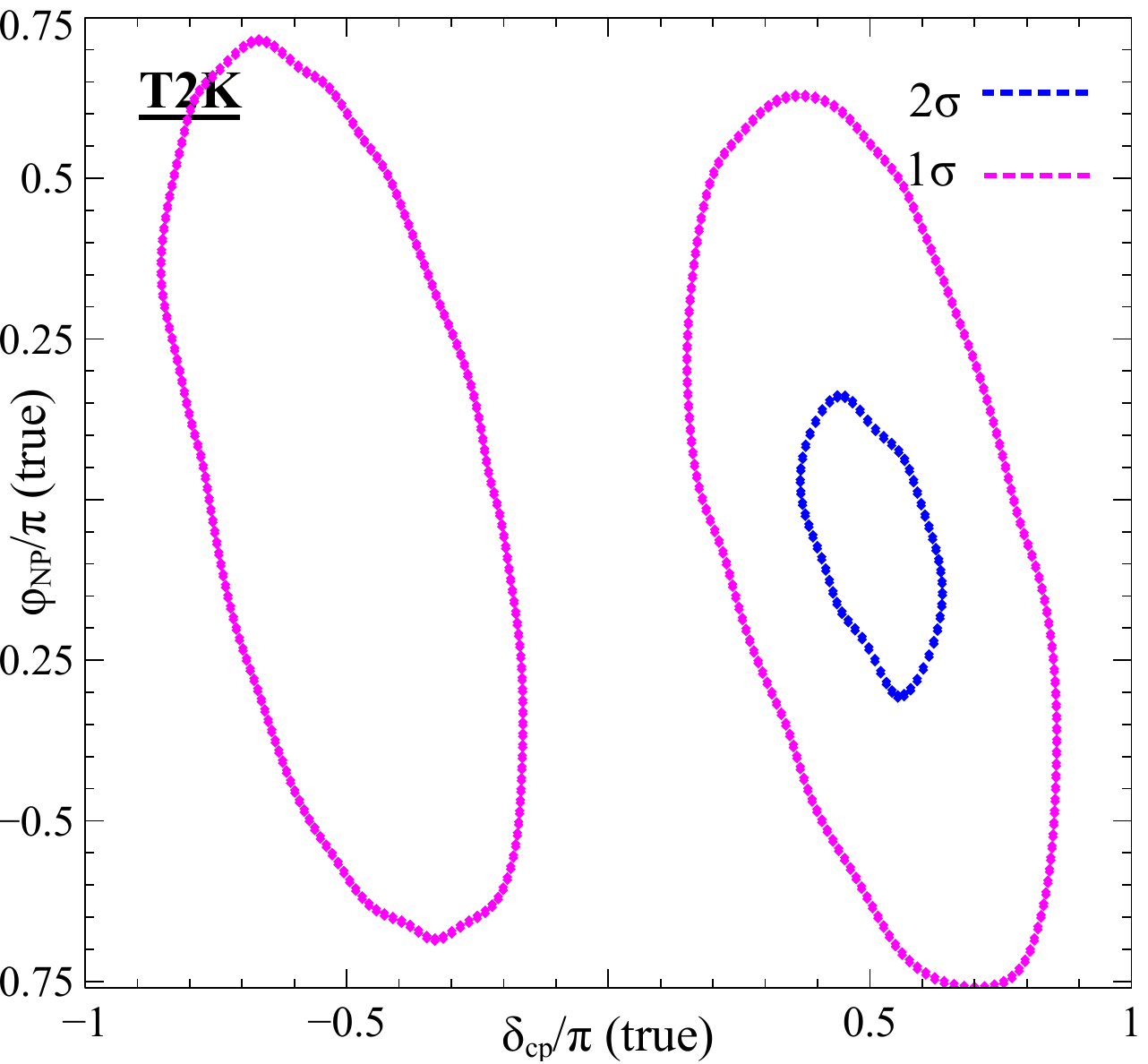}
\includegraphics[width=0.35\textwidth]{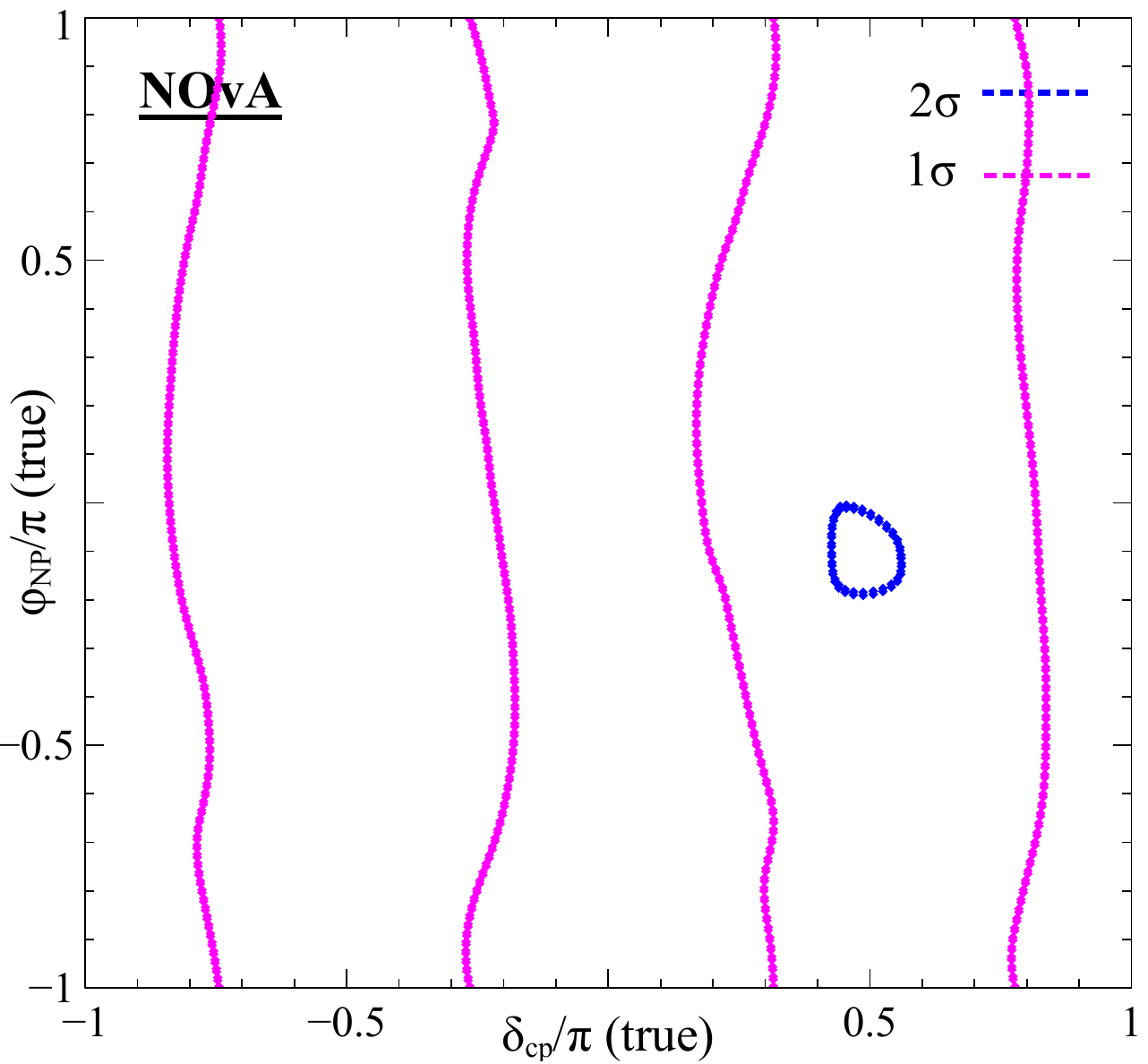}
\includegraphics[width=0.35\textwidth]{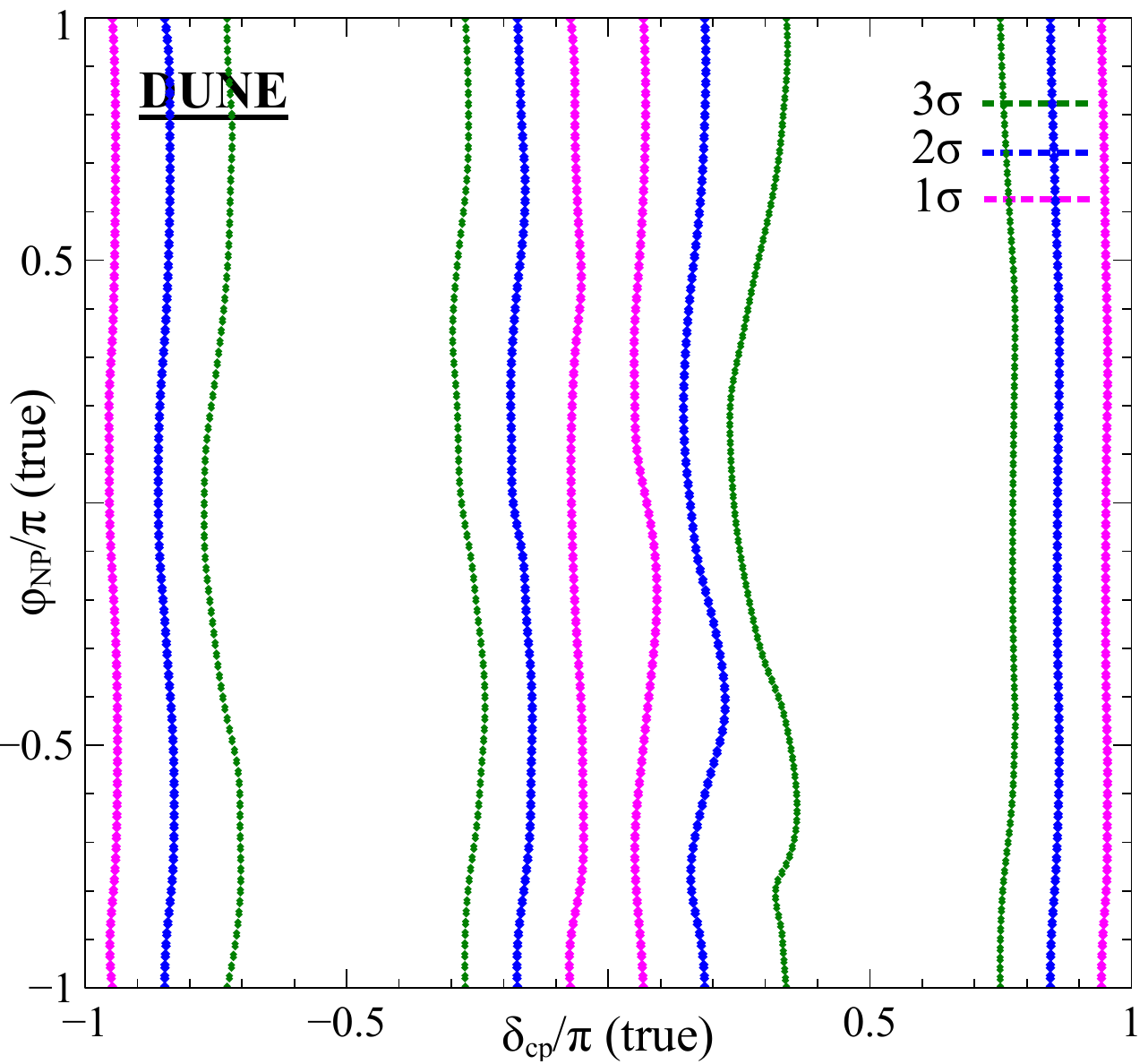}
\includegraphics[width=0.35\textwidth]{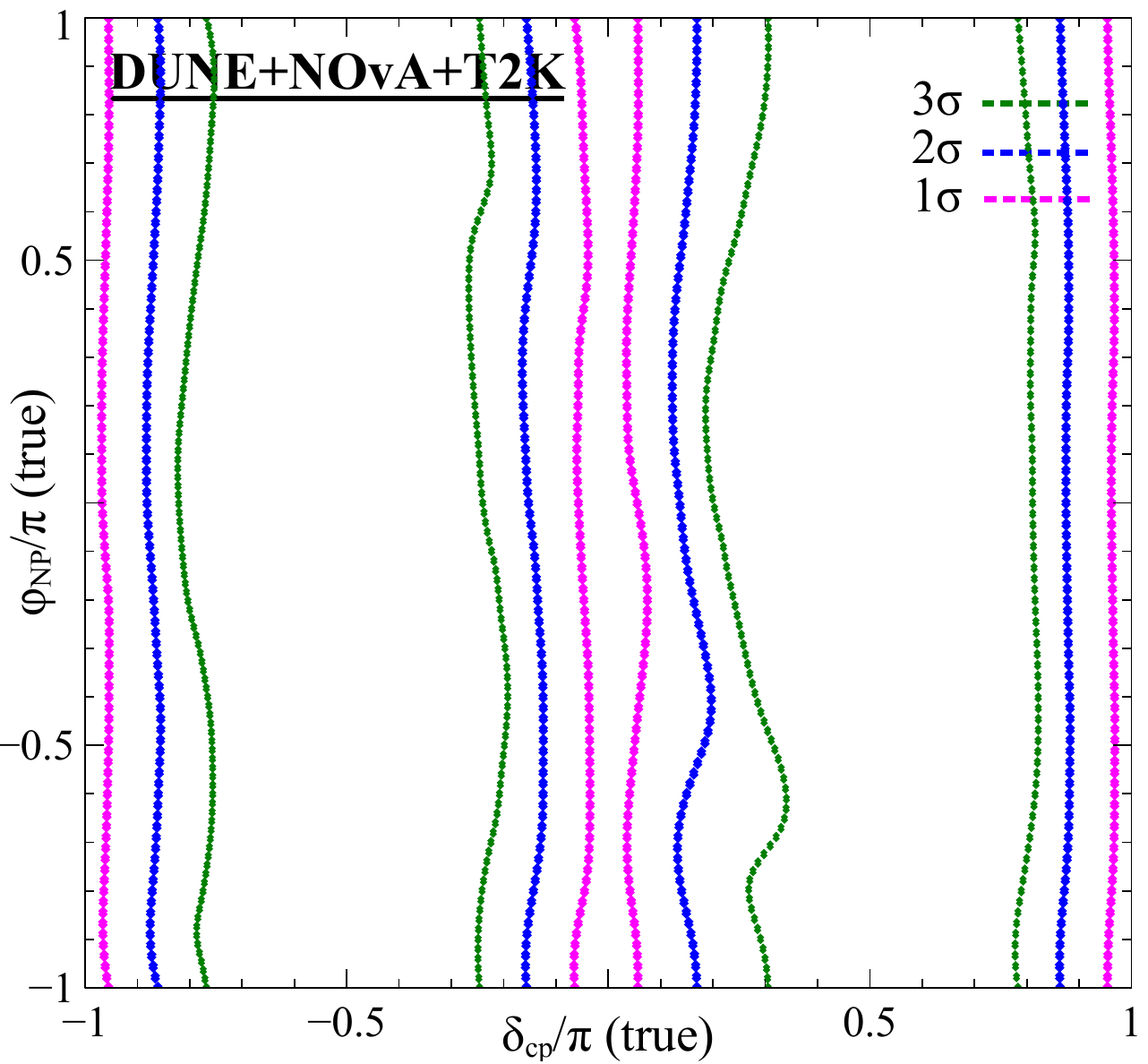}
\caption{\footnotesize{CPV discovery plots for T2K, NOvA, DUNE and their combined set ups, assuming $\delta_{CP}$ as the source of CP violation. So here in the fit, we have assumed the CP conserving values 0 and $\pi$ for $\delta_{CP}$ and marginalized $\phi_{NP}$ in its allowed range along with all other non-unitarity parameters. }}
\label{cpdis2}
\end{figure}

\subsection{CP violation discovery}
 In FIG. \ref{cpdis1} and FIG. \ref{cpdis2}, we depict the CP violation discovery potential of all the three superbeam experiments and their combinations in the presence of non-unitarity for the case of normal hierarchy. FIG. \ref{cpdis1} shows the discovery reach considering any source of CP violation for these baselines. We have shown the plots in the $\delta_{CP}(\rm true)-\phi_{NP}(\rm true)$ plane. For T2K and NO$\nu$A (upper panel of FIG. \ref{cpdis1}) the region between the blue contours corresponds to $\chi^2 \geq 2\sigma$. In these regions, any CP violation discovery is possible at or above $2\sigma$ CL. At $1\sigma$ CL, most of the space excludes CP conservation except the small region bounded by the magenta plots around $\delta_{CP} = 0^0$ and $\pm \pi$. The plots in the lower panel of FIG. \ref{cpdis1} show the CP violation discovery reach of DUNE and the combined set up. The regions outside the blue (magenta) contours are the excluded regions at $2\sigma$ ($ 1\sigma$) CL. The region outside the green lines is the excluded region at $3\sigma$ CL, and covers about 50$\%$ of the true $\delta_{CP}$ range. Combining all the experiments improves the CP violation discovery potential slightly as compared to DUNE. 
FIG. \ref{cpdis2} shows the $\delta_{CP}$ CPV discovery potential of these experiments with non-unitarity, where the CP violation originates from $\delta_{CP}$ only. As observed in the first two panels, the $\delta_{CP}$ discovery potential of T2K and NO$\nu$A is low in the presence of non-unitarity and only a $1\sigma$ discovery is possible. In DUNE and in the combined case, the discovery potential from $\delta_{CP}$ only is greater than for T2K and NO$\nu$A but less than the case of discovery from any source of CP violation. 

\section{Conclusions}

In this paper, we have attempted to analyze the sensitivity to CP violation of the long-baseline experiments NO$\nu$A, T2K and DUNE
in the presence of non-unitarity. Specifically, the effect of the extra non-unitary phase $\phi_{NP}$ is studied, which gives a degeneracy with the standard CP phase $\delta_{CP}$ and hence leads to a loss of CP violation sensitivity. Below we summarize the salient conclusions of this work:

\begin{itemize}

\item At the level of the oscillation probability $\rm P(\nu_{\mu} (\bar{\nu_{\mu}})\rightarrow \nu_e (\bar{\nu_e}))$, there is a greater variation in the probability with $\delta_{CP}$ for a given energy when the non-unitary parameters are included. If the NU phase $\phi_{NP}$ is zero, the other NU parameters have only a minor effect on the probability. The effect increases with increasing energy. The uncertainty in the probability is more when a full variation in the NU phase is included, indicating a degeneracy between $\delta_{CP}$ and $\phi_{NP}$ depicts the degeneracy between these two parameters, since for slightly separated values of energy the same probability measurement may arise from different combinations of the two phases, effectively leading to a mimicking of the standard CP phase by the NU phase. Also, an observation of CP violation may be due to either of the two phases.

\item The CP asymmetry provides further insight into the dependence of the neutrino and antineutrino oscillation probabilities on $\delta_{CP}$ and the NU parameters. It is observed that for the NH case, the standard 3-flavor CP asymmetry decreases with increasing energy, while the CP asymmetry with non-zero NU parameters but $\phi_{NP} = 0$ increases with increasing energy, indicating that at higher energies, the NU parameters, even without the NU phase, add to the uncertainty in the CP asymmetry. At energy values of about 0.6 GeV for T2K, 1.6 GeV for NO$\nu$A and 2.6 GeV for DUNE, the effect of NU parameters on the CP asymmetry is minimal and the variation due to NU parameters and $\phi_{NP}$ coincides with the asymmetry range due to $\delta_{CP}$ in the 3-flavour case. 
These values correspond to L/E $\sim$ 500 km/GeV for each experiment, which was deduced in \cite{26} to be a 'magic' value of L/E at which combining the neutrino and antineutrino channels resolve the degeneracy between $\delta_{CP}$ and $\phi_{NP}$. 
In the IH case, the presence of non-unitarity seems to nullify the CP-violating behaviour of $\delta_{CP}$ at about 1 GeV for T2K, 2.7 GeV for NO$\nu$A and 4.5 GeV for DUNE, indicating that the NU parameters, when $\phi_{NP}=0$, interact with the standard CP phase to cancel the effect of its variation on the CP asymmetry. When $\phi_{NP}$ is varied this nullifying effect no longer remains. We leave this result for the IH case as an observation which bears further study.  

\item The event rates of the experiments NO$\nu$A, DUNE and T2K reflect the behaviour of the oscillation probability when plotted as a function of reconstructed neutrino energy. The degeneracy between the standard CP phase and the NU parameters, particularly $\phi_{NP}$, shows up in the overlapping between the standard event rates and the projected event rates when non-unitarity is included.   

\item At the level of probabilities as well as event rates, the plots corresponding to $\delta_{CP}=0$ without NU and $\delta_{CP}=0, \phi_{NP}=0$ with NU show a separation between these lines, indicating that CP invariance in the $P_{\mu e}$ channel may be misinterpreted as CP violation in the three family scenario in all three experiments.

\item A $\chi^2$ analysis of the CP violation sensitivity with and without non-unitarity is done in 2 parts - CP violation irrespective of the source i.e. originating either from the standard Dirac $\delta_{CP}$ phase or from the non-unitary phase $\phi_{NP}$, and CP violation due to the standard Dirac $\delta_{CP}$ phase only in the presence of non-unitarity. In the first case, it is found that for all three experiments,  non-unitarity strongly affects the ability to measure the CP violation sensitivity, which varies on both sides of the standard 3+0 case. This is due to the interplay of the two competing effects of large parameter space and deviation from unitarity.With non-unitarity, there are 4 new parameters to be marginalised in the `fit', and the large parameter space reduces the marginalized sensitivity. Conversely, deviating from unitarity broadens the band width of sensitivity values, spreading on both sides of the 3+0 case. 
Combining the three experiments enhances the sensitivity, but the degeneracy described above is still present. If we analyze the degeneracy from the perspective of the second case i.e. CP violation due to $\delta_{CP}$ only, we find that the CP violation sensitivity decreases with non-unitarity but there is a separation between the sensitivity for the 3+0 case and the sensitivity with non-unitarity over most of the true values of $\delta_{CP}$.

Combining all the three experiments enhances the sensitivity but it is still less than the 3+0 combined sensitivity. Due to the effect of large parameter space, the CP violation sensitivity decreases uniformly compared to the 3+0 curve. 

\item CP violation discovery in the presence of non-unitarity, considering CP violation from any source, is seen to be achievable at above 1$\sigma$ for most values of true $\delta_{CP}$ and $\phi_{NP}$ and at above 2$\sigma$ over a significant range, for T2K and NO$\nu$A. The CP violation discovery reach of DUNE is above $2\sigma$ over most of the range and above 3$\sigma$ over about 50$\%$ of the range of true $\delta_{CP}$. Combining all the experiments improves the CP violation discovery potential slightly as compared to DUNE. If we consider CP violation due to $\delta_{CP}$ only, the CP violation discovery potential is reduced for all the experiments as well as their combination. 

\end{itemize}

In conclusion, this work shows that it is worthwhile to attempt CP violation sensitivity measurements at long-baseline experiments like T2K, NO$\nu$A, DUNE and their combinations. The effect of possible degeneracies of the standard CP phase with the non unitary phase and other non unitary parameters is analyzed in detail. 
Further study may be required to understand and resolve these ambiguities at other energies and baselines.
 
\begin{acknowledgments}
A special thanks to Raj Gandhi for his useful comments and suggestions in the manuscript. We acknowledge the use of the HRI cluster facility to carry out the computations. DD acknowledges support from the DAE Neutrino project at HRI. DD thanks Suprabh Prakash, Sushant K. Raut and Mehedi Masud for discussions on GLoBES. PG acknowledges local support for research at LNMIIT, Jaipur. PG also thanks Mattias Blennow and Sanjib K. Agarwalla for their comments.
\end{acknowledgments}

\end{document}